\documentclass[a4paper,11pt]{article}
\usepackage{jheppub} 
\usepackage{lineno}
\usepackage{graphics}
\usepackage{float}
\usepackage{tikz-cd} 
\usepackage{url}
\usepackage{float}
\usepackage{standalone}
\usepackage{bm}
\usepackage{hyperref}
\usepackage{cleveref}
\usepackage{subcaption}
\usepackage{longtable}
\usepackage{booktabs}
\usepackage{graphicx}
\usepackage{array}
\usepackage{paralist}
\usepackage{amsmath}
\usepackage{physics}
\usepackage{float}
\usepackage{xcolor}
\usepackage{amsfonts}
\usepackage{amssymb}
\usepackage{verbatim}
\usepackage{indentfirst}
\usepackage{tikz}
\usepackage{soul}
\usepackage{pifont}
\usepackage{forest}
\usetikzlibrary{calc}
\usetikzlibrary{arrows.meta}
\usetikzlibrary{decorations.markings}

\tikzcdset{
	arrow style=tikz,
	diagrams={>={Straight Barb[scale=0.6]}},
    every arrow/.append style={line width=0.4pt}
}
\tikzset{curve/.style={settings={#1},to path={(\tikztostart)
    .. controls ($(\tikztostart)!\pv{pos}!(\tikztotarget)!\pv{height}!270:(\tikztotarget)$)
    and ($(\tikztostart)!1-\pv{pos}!(\tikztotarget)!\pv{height}!270:(\tikztotarget)$)
    .. (\tikztotarget)\tikztonodes}},
    settings/.code={\tikzset{quiver/.cd,#1}
        \def\pv##1{\pgfkeysvalueof{/tikz/quiver/##1}}},
    quiver/.cd,pos/.initial=0.25,height/.initial=0}
\tikzset{
	solid node/.style={circle,draw,inner sep=1.2,fill=black},
	hollow node/.style={circle,draw,inner sep=1.2},
	left label/.style={above left,midway},
	right label/.style={above right,midway}
}
\usepackage{multirow} 
\usepackage{makecell}
\usepackage{longtable}

\def\diag{\mathop{\rm diag}\nolimits}

\def\tr{\mathop{\rm tr}}

\def\det{\mathop{\rm det}}
\def\beq{\begin{eqnarray}}
\def\eeq{\end{eqnarray}}
\def\beq#1\eeq{\begin{align}#1\end{align}}

\title{Seiberg dualities for quiver gauge theories}

\author[]{Yuanyuan Fang, Jing Feng, Dan Xie}
\affiliation[]{Department of Mathematics, Tsinghua University, Beijing, 100084, China}

\emailAdd{yyfang2126@gmail.com}
\emailAdd{fengj21@mails.tsinghua.edu.cn}
\emailAdd{danxie@mail.tsinghua.edu.cn}

\abstract{We study Seiberg duality for quiver gauge theories with fundamental, bifundamental, and rank-two tensor matter. The existence of a Seiberg dual description places strong constraints on the spectrum of undressed mesons, which is naturally encoded in a factorization equation for the matrix Hilbert series. We derive this factorization from a closed-form large-$N$ superconformal index for general quiver gauge theories with these matter contents. We apply the method to two-node quivers and recover known $SU$--$SU$ and $SO$--$USp$ product-group dualities. We also discuss finite-$N$ consistency conditions from anomaly and operator-spectrum matching.}

\begin{document} 
\maketitle
\flushbottom

\section{Introduction}
\label{sec:introduction}
\enlargethispage{-2\baselineskip}

\subsection{Background and motivation}

Seiberg duality~\cite{Seiberg:1994pq} is one of the most profound discoveries in the study of four-dimensional $\mathcal{N}=1$ supersymmetric gauge theories. It states that two distinct gauge theories---the ``electric'' theory and its ``magnetic'' dual---flow to the same infrared fixed point. The original duality was formulated for $SU(N_c)$ gauge theory with $N_f$ fundamental flavors. It was soon extended to $SO$ and $USp$ gauge theories with vector matter~\cite{Intriligator:1995id,Intriligator:1995ip}, to theories with adjoint or rank-two tensor matter~\cite{Kutasov:1995ss,Kutasov:1995ve,Kutasov:1995np,Intriligator:1995ff,Pouliot:1995me,Brodie:1996vx,Brodie:1996xm}, and to theories with product gauge groups~\cite{Intriligator:1995ax,Brodie:1997sz,Ahn:1997gs,Lee:1998hp,Tatar:1997xf}. A broader algebraic treatment of Seiberg duality for quiver gauge theories was developed by Berenstein and Douglas~\cite{Berenstein:2002fi}.

Superconformal indices provide another approach to the study of duality. Exact index identities have been used to organize and propose quiver-duality webs~\cite{Spiridonov:2009za,Brunner:2017lhb}. A complementary large-$N$ strategy was initiated by Kutasov and Lin~\cite{Kutasov:2014wwa}, using the large-$N$ index formula of Dolan and Osborn~\cite{Dolan:2008qi}, and was developed further by Bajc~\cite{Bajc:2019vbp} and in our previous work~\cite{Fang:2024nqy}. This strategy uses equality of the electric and magnetic large-$N$ indices to constrain the finite spectrum of undressed mesons. The present paper extends it to quiver gauge theories, for which the large-$N$ index and the resulting duality constraint take a matrix form.

\begin{figure}[H]
    \centering
\tikzset{every picture/.style={line width=0.75pt}}        
\begin{tikzpicture}[x=0.75pt,y=0.75pt,yscale=-1,xscale=1]
\draw (100,153) .. controls (100,134.77) and (114.77,120) .. (133,120) .. controls (151.23,120) and (166,134.77) .. (166,153) .. controls (166,171.23) and (151.23,186) .. (133,186) .. controls (114.77,186) and (100,171.23) .. (100,153) -- cycle ;
\draw (41,140) -- (66,140) -- (66,165) -- (41,165) -- cycle ;
\draw (201,140) -- (226,140) -- (226,165) -- (201,165) -- cycle ;
\draw (66,150) -- (100,150) ;
\draw (166,150) -- (200,150) ;
\draw (359,155) .. controls (359,136.77) and (373.77,122) .. (392,122) .. controls (410.23,122) and (425,136.77) .. (425,155) .. controls (425,173.23) and (410.23,188) .. (392,188) .. controls (373.77,188) and (359,173.23) .. (359,155) -- cycle ;
\draw (300,142) -- (325,142) -- (325,167) -- (300,167) -- cycle ;
\draw (459,142) -- (484,142) -- (484,167) -- (459,167) -- cycle ;
\draw (325,150) -- (359,150) ;
\draw (425,150) -- (459,150) ;
\draw[red] (311,141.8) .. controls (373,85.8) and (415,85.8) .. (467,142.8) ;
\draw (115,138) node [anchor=north west][inner sep=0.75pt] [align=left] {$\displaystyle \prod _{i} G_{i}$};
\draw (43,146) node [anchor=north west][inner sep=0.75pt] [align=left] {$\displaystyle n_{f}$};
\draw (203,146) node [anchor=north west][inner sep=0.75pt] [align=left] {$\displaystyle n_{f}$};
\draw (370,138) node [anchor=north west][inner sep=0.75pt] [align=left] {$\displaystyle \prod _{i} G_{i}^{\prime}$};
\draw (302,146) node [anchor=north west][inner sep=0.75pt] [align=left] {$\displaystyle n_{f}$};
\draw (462,146) node [anchor=north west][inner sep=0.75pt] [align=left] {$\displaystyle n_{f}$};
\draw (250,146) node [anchor=north west][inner sep=0.75pt] [align=left] {$\displaystyle \Rightarrow $};
\end{tikzpicture}
\caption{Schematic picture of quiver Seiberg duality. The square nodes represent flavor groups and the round nodes represent gauge groups. The label $n_f$ on a square node is schematic and may denote a class of flavor groups, not necessarily a single flavor group. The red lines connecting flavor nodes in the dual theory correspond to gauge-singlet meson fields.}
\label{fig:generalpic}
\end{figure}
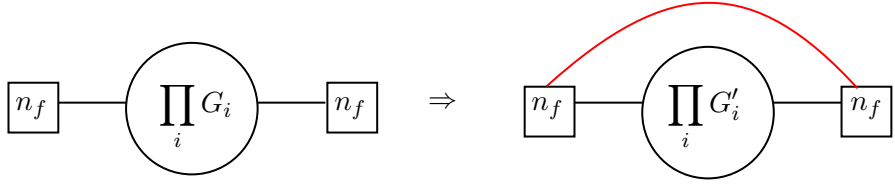

In a common approach, quiver Seiberg duality is implemented node by node, with one gauge factor dualized at a time~\cite{Berenstein:2002fi,FrancoHanany:2002}. Here we instead regard the entire gauge sector as a single unit and analyze all gauge nodes \emph{simultaneously}, as illustrated in Figure~\ref{fig:generalpic}.

\subsection{Main results}

The main results of this paper are as follows.

\begin{enumerate}
\item \textbf{Large-$N$ index formula.} We propose a new closed-form expression for the large-$N$ superconformal index of a $\prod_{i=1}^n SU(N_{ci})$ quiver gauge theory with arbitrary adjoint and bifundamental matter together with vector-like fundamental--antifundamental pairs. The formula, given in equation~\eqref{indwithfund}, takes the compact form
\begin{equation*}
\mathcal{I}(t)=\exp\Bigl(\sum_{m=1}^{\infty}\frac{1}{m}\bigl[\overline{\mathbf{g}}(t^m)^T (1-\mathbf{i}(t^m))^{-1}\mathbf{g}(t^m)-\tr\mathbf{i}(t^m)+h(t^m)\bigr]\Bigr)\prod_{m=1}^{\infty}\frac{1}{\det(1-\mathbf{i}(t^m))},
\end{equation*}
where $\mathbf{i}(t)$ is an $n\times n$ matrix encoding the gauge-sector single-letter index and $\mathbf{g}(t),\overline{\mathbf{g}}(t)$ are vectors encoding the flavor-sector contributions. The fixed-mode complex Gaussian gives a direct derivation of this formula for arbitrary $n$, while sequential large-$N$ integration over gauge nodes reproduces the same structure in explicit low-node checks and gives a conjectural formula for arbitrary $n$. The coefficient-wise Gaussian derivation is based on the Diaconis--Shahshahani moment formula~\cite{Diaconis:1994ap}: for any fixed polynomial in finitely many trace modes, its Haar integral agrees for sufficiently large $N$ with the integral of the same polynomial against the corresponding product Gaussian measure. Further details are given in Appendix~\ref{app:derive-index}. The corresponding real-Gaussian derivation for $SO/USp$ quivers is presented in Appendix~\ref{app:sosp}.

\item \textbf{Constraint equations for duality.} Imposing equality of the electric and magnetic large-$N$ indices, $\mathcal{I}_E=\mathcal{I}_M$, yields the matrix constraint
\begin{equation}\label{eq:intro-matrix-equation}
H(Q,t)-t^{\Delta+2}H(Q,t)^T=\begin{pmatrix}
\sum_{I^{11}} t^{I^{11}} & \cdots & \sum_{I^{1n}} t^{I^{1n}} \\
\vdots & \ddots & \vdots \\
\sum_{I^{n1}} t^{I^{n1}} & \cdots & \sum_{I^{nn}} t^{I^{nn}}
\end{pmatrix}.
\end{equation}
Here $H(Q,t)$ is the matrix Hilbert series of the gauge quiver, $\Delta$ is fixed by the duality pairing conditions, and the right-hand side encodes the undressed-meson $R$-charges.

This matrix equation applies to general quivers. When $H(Q,t)$ is symmetric, it reduces to
\nopagebreak
\begin{equation}\label{eq:intro-symmetric-matrix-equation}
(1-t^{\Delta+2})H(Q,t)=\begin{pmatrix}
\sum_{I^{11}} t^{I^{11}} & \cdots & \sum_{I^{1n}} t^{I^{1n}} \\
\vdots & \ddots & \vdots \\
\sum_{I^{n1}} t^{I^{n1}} & \cdots & \sum_{I^{nn}} t^{I^{nn}}
\end{pmatrix}.
\end{equation}
It gives a necessary algebraic criterion for large-$N$ index-compatible candidates within the magnetic ansatz. Its derivation and the corresponding solution strategy are presented in Section~\ref{sec:duality}.

For an invertible beta-function matrix $A$, the corresponding proposed magnetic ranks are
\begin{equation}\label{eq:intro-dual-ranks}
b_{ij}=\frac{\Delta+2}{2}(A^{-1})_{ij},
\qquad
N_{ci}^{\rm d}=\sum_{j=1}^n b_{ij}N_j-N_{ci},
\end{equation}
or, equivalently,
\begin{equation}\label{eq:intro-dual-ranks-vector}
\mathbf N_c^{\rm d}=\frac{\Delta+2}{2}A^{-1}\mathbf N-\mathbf N_c.
\end{equation}
Here $A$ is defined by the gauge-sector matter content and $R$-charges, as detailed in Section~\ref{sec:duality}.

\item \textbf{Finite-rank consistency conditions.} The proposed magnetic ranks, meson spectrum, 't Hooft anomalies, and baryon map are tested in Section~\ref{sec:anomalies}.

\item \textbf{Illustrative examples.} We apply the constraint to two symmetric two-node families. The $SU(N_{c1})\times SU(N_{c2})$ example recovers the Brodie--Hanany duality at $p=1$~\cite{Brodie:1997sz} and the general-$p$ product-group duality of Brodie~\cite{Brodie:1996vx}. We also present the two-adjoint $SO$--$USp$ family of Ahn, Oh, and Tatar~\cite{Ahn:1997gs}, whose $p=1$ member reduces to the known quartic product-group duality. The examples are developed in Section~\ref{sec:examples}.
\end{enumerate}

The paper is organized as follows. Section~\ref{sec:largen} derives the large-$N$ index for $SU$ quivers. Section~\ref{sec:duality} formulates the duality constraint, the proposed magnetic ranks, and the solution strategy. Section~\ref{sec:anomalies} discusses finite-rank anomaly conditions and baryon matching, and Section~\ref{sec:examples} applies these results to the $SU$--$SU$ and $SO$--$USp$ examples. Appendix~\ref{app:derive-index} gives the detailed complex-Gaussian derivation for $SU$ quivers, while Appendix~\ref{app:sosp} develops the corresponding real-Gaussian construction for $SO/USp$ quivers.

While preparing this manuscript for submission, we learned that Leonardo Santilli and Mohammed Akhond had independently obtained the same closed-form large-$N$ index formula using a different method~\cite{SantilliAkhond:LongQuivers}. The two works were developed independently and provide complementary derivations of the same result.

\section{Large-$N$ superconformal index for quiver gauge theories}
\label{sec:largen}

This section presents the large-$N$ superconformal index for quiver gauge theories with gauge group $\prod_{i=1}^n SU(N_{ci})$. The single-node result fixes the notation, and the general quiver formula follows in matrix form. The corresponding $SO/USp$ construction appears in Appendix~\ref{app:sosp}.

\subsection{Review: large-$N$ index of a single $SU(N_c)$ gauge theory}

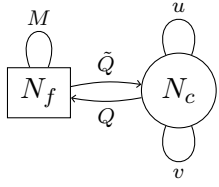
\begin{figure}[H]
    \centering
    \begin{tikzcd}
        |[draw,rectangle]| N_f\ar[loop,out=110,in=70,looseness=7,"M",no head]\ar[r,bend left=10,"\tilde Q"]&    
        |[draw,circle]| N_c \ar[l,bend left=10,"Q"]\ar[loop,out=290,in=250,looseness=5,"v",no head]\ar[loop,out=110,in=70,looseness=5,"u",no head]
    \end{tikzcd}  
    \caption{Quiver diagram for a single $SU(N_c)$ gauge theory with fundamental flavors. The round node denotes the gauge group and the square node denotes the flavor group. Adjoint chirals are drawn as loops on the gauge node, and gauge singlets as loops on the flavor node.}
    \label{fig:simple}
\end{figure}

Consider an $\mathcal{N}=1$ $SU(N_c)$ gauge theory coupled to $N_f$ fundamental chiral multiplets $Q$, $N_f$ antifundamental chiral multiplets $\tilde Q$, adjoint chiral multiplets, and gauge-singlet chiral multiplets, as shown in Figure~\ref{fig:simple}.

Following R\"omelsberger's prescription~\cite{Kinney:2005ej,Romelsberger:2005eg,Romelsberger:2007ec}, the superconformal index is obtained by taking the plethystic exponential of the single-particle index $i(t,z)$ and integrating over the Haar measure of the gauge group:
\begin{equation}\label{eq:single-index-def}
\mathcal{I}(t)=\int_{SU(N_c)} d\mu(z)\,\exp\Bigl(\sum_{m=1}^{\infty}\frac{1}{m}\,i(t^m,z^m)\Bigr).
\end{equation}
Here $z=(z_1,\dots,z_{N_c})$ are the gauge fugacities satisfying $\prod_{i=1}^{N_c}z_i=1$, and $d\mu(z)$ is the normalized Haar measure. In equation~\eqref{eq:single-index-def}, $t$ collectively denotes all fugacities other than the gauge fugacities $z$, so $t^m$ means that each non-gauge fugacity is raised to the $m$-th power. In the explicit formulas below, these fugacities are displayed separately: $t$ is normalized so that a chiral scalar of $R$-charge $R$ contributes $t^R$, $x$ is the spacetime fugacity, and $y$ denotes the flavor fugacities.

The single-particle index takes the form
\begin{equation}\label{eq:single-single-particle}
i(t,z)=f(t)\bigl(p_{N_c}(z)p_{N_c}(z^{-1})-1\bigr)+g(t)p_{N_c}(z)+\overline{g}(t)p_{N_c}(z^{-1})+h(t),
\end{equation}
where $p_{N_c}(z)=\sum_{i=1}^{N_c}z_i$ is the character of the fundamental representation. The term $p_{N_c}(z)p_{N_c}(z^{-1})-1$ is the character of the adjoint representation (the $-1$ removes the trace singlet). The coefficient functions are~\cite{Dolan:2008qi}
\begin{align}
g(t)&=\frac{t^{R_Q}-t^{2-R_Q}}{(1-tx)(1-tx^{-1})}\,\chi_{SU(N_f),\mathbf{f}},\label{eq:g-def}\\
\overline{g}(t)&=\frac{t^{R_Q}-t^{2-R_Q}}{(1-tx)(1-tx^{-1})}\,\chi_{SU(N_f),\overline{\mathbf{f}}},\label{eq:gb-def}\\
f(t)&=\frac{2t^2-t(x+x^{-1})+\sum_u(t^{R_u}-t^{2-R_u})}{(1-tx)(1-tx^{-1})},\label{eq:f-def}\\
h(t)&=\sum_M\frac{t^{R_M}-t^{2-R_M}}{(1-tx)(1-tx^{-1})}\,\chi_{SU(N_f),\mathbf{f}}\,\chi_{SU(N_f),\overline{\mathbf{f}}}.\label{eq:h-def}
\end{align}
Here $R_Q$ is the common $R$-charge of $Q$ and $\widetilde Q$, as assumed in the displayed vector-like single-node theory. The label $u$ runs over the adjoint chiral multiplets, and $R_u$ is the $R$-charge of the corresponding adjoint. Similarly, $M$ labels the gauge-singlet chiral multiplets and $R_M$ denotes their $R$-charges. For flavor fugacities $y=(y_1,\ldots,y_{N_f})$ satisfying $\prod_{a=1}^{N_f}y_a=1$, the fundamental and antifundamental characters are
\begin{equation*}
\chi_{SU(N_f),\mathbf f}(y)=\sum_{a=1}^{N_f}y_a,
\qquad
\chi_{SU(N_f),\overline{\mathbf f}}(y)=\sum_{a=1}^{N_f}y_a^{-1}.
\end{equation*}
In the large-$N$ limit, the index simplifies dramatically~\cite{Dolan:2008qi}:
\begin{equation}\label{eq:single-largeN}
\boxed{\mathcal{I}(t)=\exp\Bigl(\sum_{m=1}^{\infty}\frac{1}{m}\Bigl[\frac{g(t^m)\overline{g}(t^m)}{1-f(t^m)}-f(t^m)+h(t^m)\Bigr]\Bigr)\prod_{m=1}^{\infty}\frac{1}{1-f(t^m)}.}
\end{equation}
\subsection{General $\prod_i SU(N_{ci})$ quiver gauge theory}

Consider a quiver gauge theory with gauge group
\begin{equation*}
G=\prod_{i=1}^n SU(N_{ci}),
\end{equation*}
and a flavor symmetry group $\prod_{i=1}^n SU(N_i)_L\times SU(N_i)_R$. The matter content consists of:

\begin{itemize}
\item \textbf{Adjoint chirals} $\{X_i\}$ for each gauge node $SU(N_{ci})$. Each $X_i$ transforms in the adjoint representation $\mathbf{N_{ci}}^2-1$.
\item \textbf{Bifundamental chirals} $\{X_{ij}\}$ between gauge nodes $SU(N_{ci})$ and $SU(N_{cj})$ ($i\neq j$). Each $X_{ij}$ transforms as $(\mathbf{N_{ci}},\overline{\mathbf{N_{cj}}})$.
\item \textbf{Fundamental/antifundamental chirals} $Q_i,\tilde Q_i$ coupling the gauge node $SU(N_{ci})$ to the flavor nodes $SU(N_i)_L$ and $SU(N_i)_R$. Specifically, $Q_i$ transforms as $(\mathbf{N}_{ci},\mathbf{N}_i)$ under $SU(N_{ci})\times SU(N_i)_L$, and $\tilde Q_i$ transforms as $(\overline{\mathbf{N}}_{ci},\mathbf{N}_i)$ under $SU(N_{ci})\times SU(N_i)_R$.
\item \textbf{Gauge singlets} $\{M_{ij}\}$ between flavor groups, contributing to the singlet term $\tilde h(t)$.
\end{itemize}

The superconformal index of the quiver theory is obtained by integrating over every gauge-group factor:
\begin{equation*}
\mathcal I(t)
=\left[\prod_{i=1}^{n}\int_{SU(N_{ci})}d\mu_i(z_i)\right]
\exp\left[\sum_{m=1}^{\infty}\frac{1}{m}\,
i\bigl(t^m,z_1^m,\ldots,z_n^m\bigr)\right].
\end{equation*}
Here $d\mu_i(z_i)$ is the normalized Haar measure of the $i$-th gauge group. Thus the quiver index is a multiple integral over the product gauge group $\prod_{i=1}^n SU(N_{ci})$. The single-particle index appearing in this expression is
\begin{equation}\label{eq:general-single-particle}
\begin{aligned}
i(t,z_1,\dots,z_n)&=\sum_{i=1}^n f_i(t)\bigl(p_{N_{ci}}(z_i)p_{N_{ci}}(z_i^{-1})-1\bigr)
+\sum_{i\neq j}f_{ij}(t)\,p_{N_{ci}}(z_i)p_{N_{cj}}(z_j^{-1})\\
&\quad +\sum_{i=1}^n g_i(t)p_{N_{ci}}(z_i)+\sum_{i=1}^n \overline{g}_i(t)p_{N_{ci}}(z_i^{-1})+\tilde h(t),
\end{aligned}
\end{equation}
The coefficient functions are
\begin{align}
f_i(t)&=\frac{2t^2-t(x+x^{-1})+\sum_{X_i}(t^{R_{X_i}}-t^{2-R_{X_i}})}{(1-tx)(1-tx^{-1})},\label{eq:fi-def}\\
f_{ij}(t)&=\frac{\displaystyle\sum_{X_{ij}}t^{R_{X_{ij}}}
-\displaystyle\sum_{X_{ji}}t^{2-R_{X_{ji}}}}
{(1-tx)(1-tx^{-1})},\qquad i\neq j,\label{eq:fij-def}\\
g_i(t)&=\frac{t^{R_{Q_i}}\chi_{i,L}-t^{2-R_{Q_i}}\overline{\chi}_{i,R}}
{(1-tx)(1-tx^{-1})},\qquad
\overline{g}_i(t)=\frac{t^{R_{Q_i}}\chi_{i,R}-t^{2-R_{Q_i}}\overline{\chi}_{i,L}}
{(1-tx)(1-tx^{-1})},\label{eq:gi-def}\\
\tilde h(t)&=\sum_{i,j=1}^n h_{ij}(t),\qquad
h_{ij}(t)=\sum_{M_{ij}}\frac{t^{R_{M_{ij}}}\chi_{i,R}\chi_{j,L}
-t^{2-R_{M_{ij}}}\overline{\chi}_{i,R}\overline{\chi}_{j,L}}
{(1-tx)(1-tx^{-1})}.\label{eq:hij-def}
\end{align}
Here $\chi_{i,L}$ and $\chi_{i,R}$ are fundamental characters of $SU(N_i)_L$ and $SU(N_i)_R$, and the barred symbols are their conjugate characters. For a chiral multiplet $X_{ij}$ in $(\mathbf N_{ci},\overline{\mathbf N}_{cj})$, its scalar index letter contributes $+t^{R_{X_{ij}}}$ to $f_{ij}$, while its fermionic index letter, which transforms in the conjugate representation $(\overline{\mathbf N}_{ci},\mathbf N_{cj})$, contributes $-t^{2-R_{X_{ij}}}$ to $f_{ji}$. Thus, in~\eqref{eq:fij-def}, $f_{ij}$ receives scalar contributions from $X_{ij}$ and fermionic contributions from the independent fields $X_{ji}$. The two chiral-field sets are not identified.

\subsection{Large-$N$ index formula}

The large-$N$ index of the general quiver gauge theory can be obtained from two complementary methods.

\paragraph{Method 1: Sequential integration.} Repeated application of the single-node formula~\eqref{eq:single-largeN} integrates the gauge nodes one at a time. At the $SU(N_{c1})$ integral, the remaining gauge groups act as flavor symmetries. The resulting effective action supplies the sources for the subsequent node integrals. Explicit calculations for $n=2,3$ reproduce the determinant and inverse-kernel structure and support the general iterative form.

\paragraph{Method 2: Fixed-mode complex Gaussian.} At the $i$-th gauge node,
\begin{equation*}
z_i=(z_{i,1},\ldots,z_{i,N_{ci}}),
\qquad z_{i,\alpha}=e^{\mathrm{i}\theta_{i,\alpha}}.
\end{equation*}
The mode-$m$ variables are
\begin{equation*}
p_{i,m}=\sum_{\alpha=1}^{N_{ci}}z_{i,\alpha}^m
=\sum_{\alpha=1}^{N_{ci}}e^{\mathrm{i}m\theta_{i,\alpha}},
\qquad
p_{i,-m}=\sum_{\alpha=1}^{N_{ci}}z_{i,\alpha}^{-m}
=\sum_{\alpha=1}^{N_{ci}}e^{-\mathrm{i}m\theta_{i,\alpha}}.
\end{equation*}
Here the positive integer $m$ labels the $m$-th term in the plethystic sum, equivalently the $m$-th Fourier harmonic of the eigenvalue angles, and $p_{i,-m}=\overline{p_{i,m}}$. The Diaconis--Shahshahani moment formula applies to each individual gauge factor. Because the quiver integral uses the product Haar measure, the formula can be applied independently at every node. For each fixed coefficient in the fugacity expansion, it is enough that every gauge rank $N_{ci}$ be sufficiently large compared with the Fourier degree involving that node. At each fixed $m$, the trace modes of the $n$ gauge nodes can then be replaced by an $n$-component complex Gaussian with kernel matrix $1-\mathbf{i}(t^m)$. Evaluating this Gaussian yields the determinant and inverse-kernel structure directly, without any dependence on the order of the gauge nodes.

The basic logic is coefficient-wise. The non-gauge fugacities $t$, $x$, and $y$ grade or refine the states counted by the index, whereas the gauge fugacities $z_{i,\alpha}$ are integration variables. The Haar integrals project onto gauge-invariant states, so the gauge fugacities do not remain in the final index. We therefore organize the plethystic exponential as a formal series in $t$, keeping $x$ and the flavor fugacities $y$ as refinement parameters. For example, a scalar letter of a chiral multiplet contributes a term of the form
\begin{equation*}
t^{mR}\,\chi_{i,L}(y_i^m)\,p_{i,m}
\end{equation*}
at plethystic mode $m$. At a specified order $t^L$, this term can contribute only if $mR\leq L$. The positive $t$-degrees of the other index letters similarly bound the contributing values of $m$. Hence only finitely many plethystic modes can contribute at any fixed order in $t$, and every trace variable $p_{i,m}$ or $p_{i,-m}$ occurs to a finite power. Before the gauge integration, the coefficient of $t^L$ is therefore a finite Laurent polynomial in the gauge fugacities, equivalently a finite polynomial in finitely many trace modes. For this fixed polynomial, the Diaconis--Shahshahani moment formula~\eqref{eq:app-ds-moments} replaces the product Haar integral by the corresponding Gaussian integral once each $N_{ci}$ exceeds the Fourier degree carried by the trace variables of the $i$-th node. Thus an individual coefficient does not require taking every $N_{ci}$ literally to infinity. The full large-$N$ index is obtained by applying this finite-rank statement coefficient by coefficient. Collecting the mode-$m$ variables into $\mathbf p_m=(p_{1,m},\ldots,p_{n,m})^T$ and $\overline{\mathbf p}_m=(p_{1,-m},\ldots,p_{n,-m})^T$, the key fixed-mode integral is
\begin{align*}
&\int_{\mathbb C^n}\prod_{i=1}^n\frac{d^2p_{i,m}}{\pi m}\,
\exp\Biggl\{\frac{1}{m}\Bigl[
-\mathbf p_m^T\bigl(1-\mathbf i(t^m)\bigr)\overline{\mathbf p}_m\nonumber
+\mathbf g(t^m)^T\mathbf p_m
+\overline{\mathbf g}(t^m)^T\overline{\mathbf p}_m\Bigr]\Biggr\}\nonumber\\
&\qquad=\frac{1}{\det\bigl(1-\mathbf i(t^m)\bigr)}
\exp\Biggl[\frac{1}{m}\overline{\mathbf g}(t^m)^T\nonumber
\times\bigl(1-\mathbf i(t^m)\bigr)^{-1}\mathbf g(t^m)\Biggr].
\end{align*}
The remaining trace-independent terms at mode $m$ give $\exp\{[-\tr\mathbf i(t^m)+\tilde h(t^m)]/m\}$. Multiplying these contributions over all $m$ produces the large-$N$ index below. This infinite product is understood coefficient by coefficient, since any fixed fugacity order depends on only finitely many modes. To our knowledge, the result is a new closed-form large-$N$ index formula for $\prod_{i=1}^n SU(N_{ci})$ quiver gauge theories with arbitrary adjoint and bifundamental matter and vector-like fundamental--antifundamental pairs:

\begin{equation}\label{indwithfund}
\boxed{\begin{aligned}
\mathcal{I}(t)&\stackrel{N\to\infty}{\longrightarrow}
\exp\Bigl(\sum_{m=1}^{\infty}\frac{1}{m}\Bigl[
\overline{\mathbf{g}}(t^m)^T\bigl(1-\mathbf{i}(t^m)\bigr)^{-1}\mathbf{g}(t^m)
-\tr\mathbf{i}(t^m)+\tilde h(t^m)\Bigr]\Bigr)\\[-1pt]
&\qquad{}\times
\prod_{m=1}^{\infty}\frac{1}{\det\bigl(1-\mathbf{i}(t^m)\bigr)}.
\end{aligned}}
\end{equation}

For details of the fixed-mode Gaussian derivation, see Appendix~\ref{app:derive-index}. The corresponding real-Gaussian derivation for $SO/USp$ quivers is given in Appendix~\ref{app:sosp}.

An independent and complementary derivation of the same formula was obtained contemporaneously by Santilli and Akhond~\cite{SantilliAkhond:LongQuivers}. See the discussion in the Introduction.

Here $\mathbf{i}(t)$ is the $n\times n$ \emph{single-particle index matrix}, whose entries are
\begin{equation}\label{eq:i-matrix}
\bigl(\mathbf{i}(t)\bigr)_{ij}=
\begin{cases}
f_i(t), & i=j,\\[2pt]
f_{ij}(t), & i\neq j,
\end{cases}
\end{equation}
and $\mathbf{g}(t)=(g_1(t),\dots,g_n(t))^T$, $\overline{\mathbf{g}}(t)=(\overline{g}_1(t),\dots,\overline{g}_n(t))^T$ are the flavor-source vectors.

\subsection{Relation to the quiver matrix $M_Q(t)$}

Following~\cite{eager2012superconformal}, it is convenient to relate the matrix $1-\mathbf{i}(t)$ to the quiver adjacency data. Define the $n\times n$ matrix $M_Q(t)$ whose entries encode the $R$-charges of chiral multiplets:
\begin{equation}\label{eq:Mq-def}
\bigl(M_Q(t)\bigr)_{ij}=
\begin{cases}
\displaystyle\sum_{X_i} t^{R_{X_i}}, & i=j \quad\text{(adjoint chirals)},\\[10pt]
\displaystyle\sum_{X_{ij}} t^{R_{X_{ij}}}, & i\neq j \quad\text{(bifundamental chirals in $(\mathbf{N_{ci}},\overline{\mathbf{N_{cj}}})$)}.
\end{cases}
\end{equation}
Substituting~\eqref{eq:fi-def} and~\eqref{eq:fij-def} into~\eqref{eq:i-matrix}, the matrix $1-\mathbf{i}(t)$ can be written directly in terms of $M_Q(t)$ as
\begin{equation}\label{eq:i-to-Mq}
1-\mathbf{i}(t)=\frac{1-t^2-M_Q(t)+t^2 M_Q^T(t^{-1})}{(1-tx)(1-tx^{-1})}.
\end{equation}
The numerator in~\eqref{eq:i-to-Mq} is the inverse of the matrix Hilbert series defined below. Its entries are
\begin{equation}\label{eq:H-inverse-explicit}
\bigl(H(Q,t)^{-1}\bigr)_{ij}=
\begin{cases}
1-t^2-\displaystyle\sum_{X_i}\bigl(t^{R_{X_i}}-t^{2-R_{X_i}}\bigr), & i=j,\\[10pt]
-\displaystyle\sum_{X_{ij}}t^{R_{X_{ij}}}
+\displaystyle\sum_{X_{ji}}t^{2-R_{X_{ji}}}, & i\neq j.
\end{cases}
\end{equation}

The \emph{matrix Hilbert series} $H(Q,t)$ is defined as the matrix inverse~\cite{eager2012superconformal}
\begin{equation}\label{eq:Hilbert-series-def}
H(Q,t)=\bigl(1-t^2-M_Q(t)+t^2 M_Q^T(t^{-1})\bigr)^{-1},
\end{equation}
so that $H(Q,t)^{-1}=(1-tx)(1-tx^{-1})(1-\mathbf{i}(t))$.

\subsection{Special cases of the general formula}

\paragraph{Quivers without fundamental flavors.}
If no fundamental flavors are present, then $\mathbf{g}=\overline{\mathbf{g}}=0$, and formula~\eqref{indwithfund} becomes
\begin{equation*}
\mathcal{I}(t)=
\exp\Bigl(\sum_{m=1}^{\infty}\frac{-\tr\mathbf{i}(t^m)+\tilde h(t^m)}{m}\Bigr)
\prod_{m=1}^{\infty}\frac{1}{\det(1-\mathbf{i}(t^m))}.
\end{equation*}
For a gauge-sector quiver containing only adjoint and bifundamental matter, $\tilde h=0$. The remaining factor $\exp[-\sum_m\tr\mathbf{i}(t^m)/m]$ together with the determinant reproduces the $SU$ version of the Gadde--Rastelli--Razamat--Yan formula~\cite{Gadde:2010en}.
\paragraph{Single gauge group.}
If the quiver has a single gauge node, then $n=1$ and formula~\eqref{indwithfund} reduces to~\eqref{eq:single-largeN}. In this case, $\mathbf{i}(t)$ is the $1\times 1$ matrix $f_1(t)=f(t)$, and hence
\begin{equation*}
\det(1-\mathbf{i}(t))=1-f(t),\qquad \tr\mathbf{i}(t)=f(t),\qquad (1-\mathbf{i})^{-1}_{11}=\frac{1}{1-f(t)}.
\end{equation*}
Moreover, $\mathbf{g}$ and $\overline{\mathbf{g}}$ reduce to the scalars $g(t)$ and $\overline{g}(t)$, respectively, while $\tilde h(t)=h(t)$. Formula~\eqref{indwithfund} therefore becomes
\begin{equation*}
\mathcal{I}(t)=\exp\Bigl(\sum_{m=1}^{\infty}\frac{1}{m}\Bigl[\frac{g(t^m)\overline{g}(t^m)}{1-f(t^m)}-f(t^m)+h(t^m)\Bigr]\Bigr)\prod_{m=1}^{\infty}\frac{1}{1-f(t^m)},
\end{equation*}
which is the established single-node formula~\eqref{eq:single-largeN}~\cite{Dolan:2008qi}.

\section{Seiberg duality from the large-$N$ index}
\label{sec:duality}

Seiberg-dual electric and magnetic quivers must have equal superconformal indices. Within the magnetic ansatz below, imposing $\mathcal{I}_E=\mathcal{I}_M$ on the large-$N$ formula~\eqref{indwithfund} gives algebraic constraints on the undressed meson spectrum.

\subsection{General structure of the duality}

Consider an electric quiver gauge theory with gauge group $\prod_{i=1}^n SU(N_{ci})$ and the matter content described in Section~\ref{sec:largen}. A candidate magnetic dual theory has gauge group $\prod_{i=1}^n SU(N_{ci}^{\rm d})$ and the following properties (see Figure~\ref{fig:su-quiver-duality}):

\begin{figure}[H]
    \centering\small
    \resizebox{0.98\textwidth}{!}{
    \begin{tikzpicture}[
        gauge/.style={draw,circle,minimum size=3.0em,inner sep=1pt,align=center},
        flavor/.style={draw,rectangle,minimum width=4.8em,minimum height=1.9em,inner sep=2pt,align=center},
        field/.style={->,>=stealth},
        gaugefield/.style={->,>=stealth,shorten >=2pt,shorten <=2pt},
        meson/.style={red,thick,-},
        xlab/.style={font=\tiny,fill=white,inner sep=0.6pt},
        every node/.style={font=\scriptsize}
    ]
        \draw[blue!70,thin] (0,0) circle [radius=2.75];
        \node[gauge] (eci) at (0,1.50) {$SU(N_{ci})$};
        \node[gauge] (ecl) at (1.85,0.60) {$\cdots$};
        \node[gauge] (ecj) at (1.65,-1.45) {$SU(N_{cj})$};
        \node[gauge] (eck) at (-1.35,-1.25) {$\cdots$};
        \node[gauge] (edots) at (-1.90,0.35) {$\cdots$};

        \node[flavor] (eRi) at (0,3.60) {$SU(N_i)_R$};
        \node[flavor] (eLi) at (-4.00,1.65) {$SU(N_i)_L$};
        \node[flavor] (eLj) at (4.35,0.15) {$SU(N_j)_L$};
        \node[flavor] (eRj) at (3.35,-3.25) {$SU(N_j)_R$};

        \draw[field] (eRi) -- node[right] {$\tilde Q_i$} (eci);
        \draw[field] (eci) -- node[above left] {$Q_i$} (eLi);
        \draw[field] (ecj) -- node[above] {$Q_j$} (eLj);
        \draw[field] (eRj) -- node[right] {$\tilde Q_j$} (ecj);
        \draw[gaugefield] (eci.155) .. controls (-0.78,2.23) and (-0.26,2.40) .. node[xlab,above,pos=0.56] {$X_i$} (eci.105);
        \draw[gaugefield,bend left=6] (eci) to (ecl);
        \draw[gaugefield,bend left=6] (ecl) to (eci);
        \draw[gaugefield,bend left=6] (eci) to node[xlab,sloped,above,pos=0.30] {$X_{ij}$} (ecj);
        \draw[gaugefield,bend left=6] (ecj) to node[xlab,sloped,below,pos=0.30] {$X_{ji}$} (eci);
        \draw[gaugefield,bend left=6] (eci) to (eck);
        \draw[gaugefield,bend left=6] (eck) to (eci);
        \draw[gaugefield,bend left=6] (ecl) to (ecj);
        \draw[gaugefield,bend left=6] (ecj) to (ecl);
        \draw[gaugefield,bend left=6] (ecl) to (eck);
        \draw[gaugefield,bend left=6] (eck) to (ecl);
        \draw[gaugefield,bend left=6] (eck) to (ecj);
        \draw[gaugefield,bend left=6] (ecj) to (eck);
        \draw[gaugefield,bend left=6] (edots) to (eci);
        \draw[gaugefield,bend left=6] (eci) to (edots);
        \draw[gaugefield,bend left=6] (eck) to (edots);
        \draw[gaugefield,bend left=6] (edots) to (eck);
        \node at (5.45,0) {$\stackrel{\text{Dual}}{\Longrightarrow}$};

        \draw[blue!70,thin] (10.80,0) circle [radius=2.75];
        \node[gauge] (mci) at (10.80,1.50) {$SU(N_{ci}^{\rm d})$};
        \node[gauge] (mcl) at (12.65,0.60) {$\cdots$};
        \node[gauge] (mcj) at (12.45,-1.45) {$SU(N_{cj}^{\rm d})$};
        \node[gauge] (mck) at (9.45,-1.25) {$\cdots$};
        \node[gauge] (mdots) at (8.90,0.35) {$\cdots$};

        \node[flavor] (mRi) at (10.80,3.60) {$SU(N_i)_R$};
        \node[flavor] (mLi) at (6.80,1.65) {$SU(N_i)_L$};
        \node[flavor] (mLj) at (15.15,0.15) {$SU(N_j)_L$};
        \node[flavor] (mRj) at (14.15,-3.25) {$SU(N_j)_R$};

        \draw[field] (mci) -- node[right] {$q_i$} (mRi);
        \draw[field] (mLi) -- node[above left] {$\tilde q_i$} (mci);
        \draw[field] (mLj) -- node[above] {$\tilde q_j$} (mcj);
        \draw[field] (mcj) -- node[right] {$q_j$} (mRj);
        \draw[gaugefield] (mci.155) .. controls (10.02,2.23) and (10.54,2.40) .. node[xlab,above,pos=0.56] {$X_i^*$} (mci.105);
        \draw[gaugefield,bend left=6] (mcl) to (mci);
        \draw[gaugefield,bend left=6] (mci) to (mcl);
        \draw[gaugefield,bend left=6] (mcj) to node[xlab,sloped,below,pos=0.30] {$X_{ij}^*$} (mci);
        \draw[gaugefield,bend left=6] (mci) to node[xlab,sloped,above,pos=0.30] {$X_{ji}^*$} (mcj);
        \draw[gaugefield,bend left=6] (mck) to (mci);
        \draw[gaugefield,bend left=6] (mci) to (mck);
        \draw[gaugefield,bend left=6] (mcj) to (mcl);
        \draw[gaugefield,bend left=6] (mcl) to (mcj);
        \draw[gaugefield,bend left=6] (mck) to (mcl);
        \draw[gaugefield,bend left=6] (mcl) to (mck);
        \draw[gaugefield,bend left=6] (mcj) to (mck);
        \draw[gaugefield,bend left=6] (mck) to (mcj);
        \draw[gaugefield,bend left=6] (mci) to (mdots);
        \draw[gaugefield,bend left=6] (mdots) to (mci);
        \draw[gaugefield,bend left=6] (mdots) to (mck);
        \draw[gaugefield,bend left=6] (mck) to (mdots);
        \draw[meson] (mRi.north east) .. controls (12.00,4.80) and (14.65,3.45) .. node[pos=0.45,above] {$M_{I^{ij}}$} (mLj.north east);
        \draw[meson] (mLi.south east) .. controls (7.45,-0.25) and (7.45,-3.90) .. (9.25,-3.90)
            .. controls (11.80,-3.90) and (13.10,-3.55) .. node[pos=0.58,above] {$M_{I^{ji}}$} (mRj.south west);
    \end{tikzpicture}}
    \caption{Schematic picture of the $SU$ quiver duality. The square nodes denote flavor groups and the round nodes denote gauge groups. The blue circles enclose the gauge-sector part of the quiver. The round nodes labeled by $\cdots$ represent arbitrary additional gauge nodes, each of which may have its own adjoint chiral multiplets and attached flavor groups omitted from the drawing. The flavor groups attached to $SU(N_{ci})$ are $SU(N_i)_L$ and $SU(N_i)_R$, where $i=1,2,\ldots,n$. The loops on $SU(N_{ci})$ and $SU(N_{ci}^{\rm d})$ denote adjoint chiral multiplets $X_i$ and $X_i^*$, respectively. The displayed opposite arrows illustrate the two possible orientations: $X_{ij}$ and $X_{ji}$ are independent field sets, and either orientation may be absent or may have a different multiplicity and $R$-charge spectrum. Each magnetic arrow is oriented oppositely to its electric partner. The red lines in the magnetic quiver denote elementary meson singlets, including $M_{I^{ij}}$ and $M_{I^{ji}}$.}
    \label{fig:su-quiver-duality}
\end{figure}
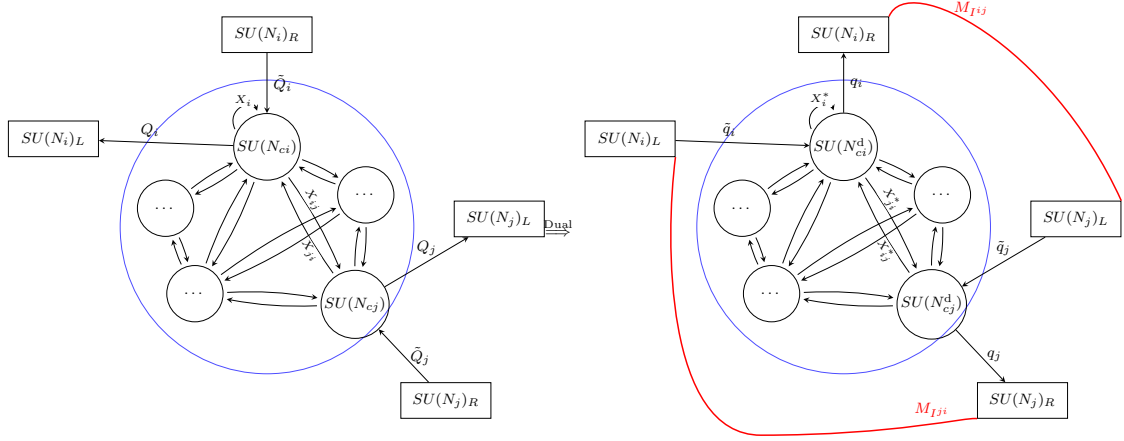

\begin{enumerate}
\item \textbf{Field replacement.} Every chiral field of the electric theory is replaced by a dual field in the conjugate gauge representation:
\begin{equation*}
Q_i\to \tilde q_i,\quad \tilde Q_i\to q_i,\quad X_i\to X_i^*,\quad X_{ij}\to X_{ij}^*.
\end{equation*}
Here $\tilde q_i$ transforms as $(\overline{\mathbf{N}}_{ci}^{\rm d},\overline{\mathbf{N}}_i)$ under $SU(N_{ci}^{\rm d})\times SU(N_i)_L$, while $q_i$ transforms as $(\mathbf{N}_{ci}^{\rm d},\overline{\mathbf{N}}_i)$ under $SU(N_{ci}^{\rm d})\times SU(N_i)_R$. Similarly, $X_{ij}$ transforms as $(\mathbf{N}_{ci},\overline{\mathbf{N}}_{cj})$ under $SU(N_{ci})\times SU(N_{cj})$, while $X_{ij}^*$ transforms as $(\overline{\mathbf{N}}_{ci}^{\rm d},\mathbf{N}_{cj}^{\rm d})$ under $SU(N_{ci}^{\rm d})\times SU(N_{cj}^{\rm d})$. This replacement is made independently for every oriented bifundamental. No equality between the $X_{ij}$ and $X_{ji}$ field sets is assumed.
The flavor groups $\prod_i SU(N_i)_L\times SU(N_i)_R$ are unchanged.

\item \textbf{Gauge singlets.} For each oriented path from a flavor node $SU(N_j)_L$ to a flavor node $SU(N_i)_R$ through the gauge quiver, there exists a dressed meson operator
\begin{equation*}
M_{I^{ij}}=\tilde Q_i\,U_{I^{ij}}\,Q_j,
\end{equation*}
where $U_{I^{ij}}$ is an undressed meson (a word in the adjoint and bifundamental chirals). This meson transforms as $(\mathbf N_i,\mathbf N_j)$ under $SU(N_i)_R\times SU(N_j)_L$. The set $\{U_{I^{ij}}\}$ must obey a quantum truncation: only finitely many such operators are independent. The total number of mesons in the $(i,j)$ channel is denoted $a_{ij}$. These electric mesons appear as elementary gauge singlets $[M_{I^{ij}}]$ in the magnetic theory. We assume that the corresponding electric and magnetic gauge-sector words have the same finite set of $R$-charges, including multiplicities,
\begin{equation*}
\bigl\{I^{ij}\bigr\}=\bigl\{I^{ij\,\prime}\bigr\}.
\end{equation*}

\item \textbf{Superpotential.} In $W_E$, each dressed meson is replaced by the corresponding elementary magnetic singlet $[M_{I^{ij}}]$. New cubic couplings are then added:
\begin{equation}\label{eq:deltaw}
\Delta W=\sum_{i,j=1}^n\sum_{I^{ij}}[M_{I^{ij}}]\,M_{I^{ij\,\prime}}',
\end{equation}
where $M_{I^{ij\,\prime}}'$ is the dressed meson built from the dual fields whose undressed word has charge $I^{ij\,\prime}$. The F-term equation of the elementary singlet sets the corresponding magnetic composite $M_{I^{ij\,\prime}}'$ to zero in the chiral ring. In this sense, the superpotential coupling lifts, or integrates out, these composite chiral operators, preventing them from being counted as additional independent mesons in the magnetic theory.

\item \textbf{Pairing condition.} The new superpotential terms require that the $R$-charges of the undressed mesons be paired:
\begin{equation}\label{eq:pairing}
I^{ij}+I^{ij\,\prime}=\Delta,\qquad i,j=1,\dots,n,
\end{equation}
for some constant $\Delta$. The equality of the two charge sets does not mean that each charge is paired with an identical charge. Equation~\eqref{eq:pairing} specifies the pairing and requires the common set to be invariant, including multiplicities, under $I\mapsto\Delta-I$. Here $I^{ij}$ and $I^{ij\,\prime}$ are the $R$-charges of the electric and magnetic gauge-sector words, respectively. They do not include the charges of the quarks at the two ends of the path. The elementary singlet $[M_{I^{ij}}]$ inherits the electric meson charge $R_{Q_i}+R_{Q_j}+I^{ij}$, and the magnetic dressed meson has charge $R_{q_i}+R_{q_j}+I^{ij\,\prime}$. Requiring every term in~\eqref{eq:deltaw} to have $R$-charge two gives
\begin{equation*}
\bigl(R_{Q_i}+R_{Q_j}+I^{ij}\bigr)
+\bigl(R_{q_i}+R_{q_j}+I^{ij\,\prime}\bigr)=2.
\end{equation*}
Using~\eqref{eq:pairing}, this becomes
\begin{equation*}
\bigl(R_{Q_i}+R_{q_i}\bigr)+\bigl(R_{Q_j}+R_{q_j}\bigr)+\Delta=2.
\end{equation*}
In particular, the diagonal channel $i=j$ gives the quark pairing condition
\begin{equation}\label{eq:quark-pairing}
2R_{Q_i}+2R_{q_i}+\Delta=2,\qquad i=1,\dots,n.
\end{equation}
\end{enumerate}

This list is an ansatz for the magnetic theory. In particular, we assume the same gauge-sector quiver structure with conjugate gauge representations, elementary singlets identified with finitely many mesons, and the pairing condition above. Magnetic descriptions with extra singlets, different gauge-sector $R$-charges, or different matter content are not classified by the criterion derived here.

\subsection{Beta functions and dual gauge ranks}

For a general quiver, gauge-anomaly cancellation must be imposed before the NSVZ conditions. Let $n_{ij}$ be the number of chirals in the set $X_{ij}$. Since $Q_i$ and $\tilde Q_i$ have opposite gauge chirality, their contributions cancel, and the electric $SU(N_{ci})^3$ gauge anomaly requires
\begin{equation}\label{eq:gauge-anomaly-balance}
\sum_{j\neq i}(n_{ij}-n_{ji})N_{cj}=0,
\qquad i=1,\dots,n.
\end{equation}
The conjugate magnetic arrows must satisfy the analogous condition with $N_{cj}$ replaced by $N_{cj}^{\rm d}$. These weighted balance conditions allow unequal arrow multiplicities in the two orientations, but exclude anomalous matter assignments.

The vanishing of the NSVZ beta functions for the electric theory imposes $n$ constraints:
\begin{equation}\label{eq:electric-beta}
\left(1+\sum_{X_i}(R_{X_i}-1)\right)N_{ci}
+\frac{1}{2}\sum_{j\neq i}N_{cj}
\left[\sum_{X_{ij}}(R_{X_{ij}}-1)+\sum_{X_{ji}}(R_{X_{ji}}-1)\right]
+N_i(R_{Q_i}-1)=0,
\end{equation}
for $i=1,\dots,n$. Introducing the $n\times n$ matrix $A$ with entries
\begin{equation}\label{eq:A-matrix}
A_{ij}=
\begin{cases}
1+\sum_{X_i}(R_{X_i}-1), & i=j,\\[4pt]
\frac{1}{2}\left[\displaystyle\sum_{X_{ij}}(R_{X_{ij}}-1)
+\displaystyle\sum_{X_{ji}}(R_{X_{ji}}-1)\right], & i\neq j,
\end{cases}
\end{equation}
the beta function conditions can be written compactly as
\begin{equation}\label{eq:beta-matrix}
\sum_{j=1}^n A_{ij}N_{cj}+N_i(R_{Q_i}-1)=0,\qquad i=1,\dots,n.
\end{equation}
Solving for the quark $R$-charges gives
\begin{equation}\label{eq:RQi}
R_{Q_i}=1-\frac{1}{N_i}\sum_{j=1}^n A_{ij}N_{cj}.
\end{equation}
The two sums in the off-diagonal entry are independent, so the definition accommodates unequal arrow multiplicities in the two orientations. The matrix $A$ is nevertheless symmetric because a bifundamental contributes to the beta functions of both endpoint gauge nodes. We assume that $A$ is invertible when solving for the magnetic ranks.

Similarly, the vanishing of the magnetic beta functions yields
\begin{equation}\label{eq:magnetic-beta}
\sum_{j=1}^n A_{ij}N_{cj}^{\rm d}+N_i(R_{q_i}-1)=0,\qquad i=1,\dots,n.
\end{equation}
Using the pairing condition~\eqref{eq:quark-pairing} to express $R_{q_i}$ in terms of $R_{Q_i}$, define the symmetric beta-function coefficient matrix
\begin{equation}\label{eq:b-matrix}
b_{ij}:=\frac{\Delta+2}{2}(A^{-1})_{ij}.
\end{equation}
The magnetic beta functions then give
\begin{equation}\label{eq:dual-ranks}
N_{ci}^{\rm d}=\sum_{j=1}^n b_{ij}N_j-N_{ci}.
\end{equation}
In vector notation,
\begin{equation}\label{eq:dual-ranks-vector}
\boxed{\mathbf{N_c^{\rm d}}=\frac{\Delta+2}{2}A^{-1}\mathbf{N}-\mathbf{N_c},}
\end{equation}
where $\mathbf{N_c}=(N_{c1},\dots,N_{cn})^T$ and $\mathbf{N}=(N_1,\dots,N_n)^T$.

It is important not to identify $b_{ij}$ with a meson multiplicity at this stage. We reserve $a_{ij}=|\mathcal I^{ij}|$ for the number of mesons read from the $(i,j)$ entry of the index constraint. The matrix $b$ controls the proposed magnetic ranks, whereas $a_{ij}$ counts mesons in the $(i,j)$ flavor channel. For arbitrary independent flavor ranks, finite-rank cubic flavor-anomaly matching requires $a_{ij}=a_{ji}=b_{ij}$, as discussed in Section~\ref{sec:anomalies}.

In summary, the matrix $A$ encodes the quiver matter content and its gauge-sector $R$-charges. Once this electric data is fixed, the remaining data of the duality ansatz are the constant $\Delta$ and the finite sets of undressed-meson charges $\{I^{ij}\}$. The constant $\Delta$ determines the quark pairing through~\eqref{eq:quark-pairing} and enters the proposed magnetic ranks in~\eqref{eq:dual-ranks-vector}, while each set $\{I^{ij}\}$ specifies the elementary meson singlets in the corresponding flavor channel. The charge sets must admit the complementary pairing $I^{ij}+I^{ij\,\prime}=\Delta$. Their allowed values are determined below by imposing equality of the electric and magnetic large-$N$ indices.

\subsection{Equality of large-$N$ indices and the matrix constraint}

Equality of the electric and magnetic large-$N$ indices requires
\begin{equation*}
\mathcal{I}_E=\mathcal{I}_M.
\end{equation*}
We assume that every magnetic gauge-sector field has the same $R$-charge as its electric partner and lies in the conjugate gauge representation. For a general quiver this gives
\begin{equation}\label{eq:kernel-transpose}
\mathbf i_M(t)=\mathbf i_E(t)^T,\qquad
K_M(t)=K_E(t)^T,\qquad K_E(t):=1-\mathbf i_E(t).
\end{equation}
Thus $\det K_M=\det K_E$, so the determinant factors cancel without requiring $K_E$ to be symmetric. The remaining equality involves the bilinear flavor-source terms $\overline{\mathbf g}^{\,T}K^{-1}\mathbf g$ in~\eqref{indwithfund} and the elementary-singlet terms $\tilde h^E$ and $\tilde h^M$ defined in~\eqref{eq:hij-def}. In particular, the elementary magnetic mesons contribute through $\tilde h^M$. The condition $\mathcal I_E=\mathcal I_M$ can therefore be simplified as
\begin{equation}\label{eq:index-equality-reduced}
\begin{aligned}
0
&=(\overline{\mathbf g}^{E})^T K_E^{-1}\mathbf g^E
-(\overline{\mathbf g}^{M})^T K_M^{-1}\mathbf g^M
+\tilde h^E-\tilde h^M\\
&=(\overline{\mathbf g}^{E})^T K_E^{-1}\mathbf g^E
-(\mathbf g^M)^T K_E^{-1}\overline{\mathbf g}^M
+\tilde h^E-\tilde h^M.
\end{aligned}
\end{equation}
The second line follows from $K_M=K_E^T$ and the fact that each bilinear source term is a scalar. The plethystic sum $\sum_m\frac{1}{m}$ is understood. If a proposed duality changes the gauge-sector $R$-charges or adds gauge-sector fields, even this transpose relation may fail and the following constraint must be modified.

We take the electric theory to have no elementary gauge singlets, so $\tilde h^E=0$. Writing $D=(1-tx)(1-tx^{-1})$, the magnetic quark sources, with the representations in Section~\ref{sec:duality}, are
\begin{equation}\label{eq:magnetic-sources}
g_i^M=\frac{t^{R_{q_i}}\overline\chi_{i,R}-t^{2-R_{q_i}}\chi_{i,L}}{D},
\qquad
\overline g_i^M=\frac{t^{R_{q_i}}\overline\chi_{i,L}-t^{2-R_{q_i}}\chi_{i,R}}{D},
\end{equation}
The magnetic theory also contains elementary meson singlets $M_{I^{ij}}$ in the character $\chi_{i,R}\chi_{j,L}$. Matching this ordered flavor character in~\eqref{eq:index-equality-reduced} gives
\begin{equation}\label{eq:hij-equation}
\frac{t^{R_{Q_i}+R_{Q_j}}}{D}\sum_{I^{ij}}t^{I^{ij}}
=\frac{t^{R_{Q_i}+R_{Q_j}}}{D^2}(K_E^{-1})_{ij}
-\frac{t^{4-R_{q_i}-R_{q_j}}}{D^2}(K_E^{-1})_{ji}.
\end{equation}
The first term comes from the electric scalar letters in $\overline g_i^Eg_j^E$, while the second comes from the magnetic fermionic letters in $g_j^M\overline g_i^M$. The $(i,j)$ and $(j,i)$ equations are therefore distinct when the multiplicities or $R$-charge spectra of arrows in the two orientations differ. To express this equality directly in terms of the matrix Hilbert series, we use $K_E^{-1}=D\,H(Q,t)$ and
\[
4-R_{q_i}-R_{q_j}-R_{Q_i}-R_{Q_j}=\Delta+2,
\]
Then equation~\eqref{eq:hij-equation} gives
\begin{equation}\label{eq:index-constraint-scalar}
H_{ij}(Q,t)-t^{\Delta+2}H_{ji}(Q,t)
=\sum_{I^{ij}}t^{I^{ij}},
\qquad i,j=1,\dots,n.
\end{equation}
No symmetry of the quiver matrix $M_Q(t)$ defined in~\eqref{eq:Mq-def}, or of the matrix Hilbert series $H(Q,t)$ defined in~\eqref{eq:Hilbert-series-def}, is required. Equivalently, the matrix form of the large-$N$ index constraint is

\begin{equation}\label{eq:dual-matrix-form}
\boxed{H(Q,t)-t^{\Delta+2}H(Q,t)^T=\begin{pmatrix}
\displaystyle\sum_{I^{11}}t^{I^{11}} & \cdots & \displaystyle\sum_{I^{1n}}t^{I^{1n}} \\[8pt]
\vdots & \ddots & \vdots \\[4pt]
\displaystyle\sum_{I^{n1}}t^{I^{n1}} & \cdots & \displaystyle\sum_{I^{nn}}t^{I^{nn}}
\end{pmatrix}.}
\end{equation}

If $H(Q,t)$ is symmetric, equation~\eqref{eq:dual-matrix-form} reduces to
\[
(1-t^{\Delta+2})H(Q,t)=
\begin{pmatrix}
\displaystyle\sum_{I^{11}}t^{I^{11}} & \cdots & \displaystyle\sum_{I^{1n}}t^{I^{1n}} \\[8pt]
\vdots & \ddots & \vdots \\[4pt]
\displaystyle\sum_{I^{n1}}t^{I^{n1}} & \cdots & \displaystyle\sum_{I^{nn}}t^{I^{nn}}
\end{pmatrix}.
\]

The right-hand side of~\eqref{eq:dual-matrix-form} represents the candidate spectrum of undressed mesons, with each monomial $t^{I^{ij}}$ corresponding to an undressed meson of $R$-charge $I^{ij}$. Within the magnetic ansatz, every matrix entry must therefore be a finite polynomial with non-negative integer coefficients, reflecting the truncation to finitely many independent mesons.

\subsection{Solution strategy}

To turn the large-$N$ index constraint into a practical test for candidate dualities, we first clear the fractional powers and recall the single-node factorization before stating the general quiver procedure. For rational exponents, choose a positive integer $\ell$ such that $\ell\alpha\in\mathbb Z$ for every rational exponent $\alpha$ appearing in the constraint, and set
\[
t=y^\ell,\qquad L=\ell(\Delta+2)\in\mathbb Z_{>0}.
\]

For a single gauge group with adjoint chirals $X_a$ of $R$-charges $R_a$, one has
\[
H(Q,t)^{-1}=1-t^2-\sum_a\bigl(t^{R_a}-t^{2-R_a}\bigr).
\]
The $n=1$ specialization of~\eqref{eq:dual-matrix-form} is therefore equivalent to the scalar index constraint~\cite{Kutasov:2014wwa}
\begin{equation}\label{eq:single-node-polynomial-constraint}
t^{\Delta+2}-1
=\left[-1+t^2+\sum_a\bigl(t^{R_a}-t^{2-R_a}\bigr)\right]
\sum_I t^I .
\end{equation}
With
\[
\Phi_+(y)=\sum_I y^{\ell I},
\qquad
\Phi_-(y)=-1+y^{2\ell}
+\sum_a\left(y^{\ell R_a}-y^{\ell(2-R_a)}\right),
\]
equation~\eqref{eq:single-node-polynomial-constraint} takes the cyclotomic factorization form~\cite{Bajc:2019vbp,Fang:2024nqy}
\begin{equation}\label{eq:single-node-cyclotomic-factorization}
y^L-1=\Phi_-(y)\Phi_+(y).
\end{equation}
For a general quiver, the same change of variables turns~\eqref{eq:dual-matrix-form} into a matrix polynomiality problem. The complete procedure is as follows.

\begin{enumerate}
\item Given a quiver $Q$ (the gauge sector without flavors), compute the matrix Hilbert series $H(Q,t)$ using~\eqref{eq:Hilbert-series-def}.
\item Choose $\ell$ as above so that every exponent in $H(Q,t)$ becomes an integer after setting $t=y^\ell$. Treat $L=\ell(\Delta+2)$ as an unknown positive integer. The tests below determine the allowed values of $L$. Once $L$ is fixed, $\Delta=L/\ell-2$.
\item For each proposed $L$, form the matrix
\[
\mathcal P_L(y)
:=H(Q,t=y^\ell)-y^L H(Q,t=y^\ell)^T.
\]
The proposed value passes the large-$N$ polynomiality test only if, after combining the terms in each entry and canceling all common numerator-denominator factors, every entry of $\mathcal P_L(y)$ is a finite polynomial with non-negative integer coefficients. When $H(Q,t)$ is nonsymmetric, the denominators of $H_{ij}$ and $H_{ji}$ cannot be analyzed separately. For each proposed $L$, one must explicitly form and simplify every entry $H_{ij}-y^L H_{ji}$, and then check its polynomiality and coefficient positivity.
\item If $H(Q,t)$ is symmetric, then $\mathcal P_L(y)=(1-y^L)H(Q,t=y^\ell)$. Reduce each entry to
\[
H_{ij}(Q,t=y^\ell)=\frac{p_{ij}(y)}{q_{ij}(y)},
\]
where $p_{ij}$ and $q_{ij}$ have no common factors. Polynomiality requires all factors of $q_{ij}(y)$ to be canceled by $1-y^L$. The numerator cannot contribute to this cancellation because the fraction is reduced. The standard factorization
\[
y^L-1=\prod_{\substack{d>0\\ d\text{ divides }L}}\Phi_d(y)
\]
shows that $1-y^L$ has only cyclotomic factors, each with multiplicity one. Hence every reduced denominator must have the form
\[
q_{ij}(y)=\prod_d\Phi_d(y)^{m_{ij,d}},
\qquad m_{ij,d}\in\{0,1\}.
\]
If any $q_{ij}$ contains a non-cyclotomic factor or a repeated cyclotomic factor, no value of $L$ can pass the denominator test.
\item When all reduced denominators pass this test, collect the cyclotomic orders
\[
\mathcal N=\{d:\Phi_d(y)\text{ appears in some }q_{ij}(y)\}.
\]
Every admissible $L$ must be divisible by each $d\in\mathcal N$, so the smallest value allowed by the denominator test is
\[
L_{\min}=\operatorname{lcm}(\mathcal N).
\]
If $\mathcal N$ is empty, this step imposes no divisibility condition and we set $L_{\min}=1$. The denominator test is only necessary: the values not excluded by it are positive multiples of $L_{\min}$, and each proposed value must still be tested using the full matrix $\mathcal P_L(y)$, including coefficient positivity.
\item For every $L$ that passes the full test, read a monomial $y^a$ in $(\mathcal P_L)_{ij}$ as an undressed meson of $R$-charge $I^{ij}=a/\ell$, and let $a_{ij}$, the number of mesons in that channel, be the sum of the coefficients in that entry. Next, one can compute the proposed dual ranks from~\eqref{eq:dual-ranks}.
\end{enumerate}

When $H(Q,t)$ is nonsymmetric, the entries $H_{ij}$ and $H_{ji}$ are coupled, and there is no channel-by-channel cyclotomic criterion analogous to the symmetric case. For a specified candidate value of $\Delta$, one must form
\[
H(Q,t)-t^{\Delta+2}H(Q,t)^T
\]
and check that every entry is a finite polynomial with non-negative integer coefficients. In this paper, the explicit cyclotomic analysis is applied to the symmetric kernels appearing in the examples.

\section{Finite-rank anomaly conditions and baryon spectrum}
\label{sec:anomalies}

The large-$N$ constraint~\eqref{eq:dual-matrix-form} determines a candidate meson spectrum, whereas the proposed magnetic ranks are obtained separately from~\eqref{eq:dual-ranks}. Accordingly, the meson multiplicities $a_{ij}$ extracted from the index and the coefficients $b_{ij}$ entering the magnetic-rank formula need not coincide a priori. We now derive the relations imposed by finite-rank anomaly matching and then examine the baryon spectrum. Together, these provide independent consistency tests of the proposed finite-rank duality.

\subsection{Global symmetry}

The nonabelian flavor symmetry is $\prod_i SU(N_i)_L\times SU(N_i)_R$. To discuss baryonic symmetries, it is useful to introduce baryon charges $B_i$. Only anomaly-free linear combinations of these baryon charges define physical baryonic symmetries. An individual $B_i$ need not be a global symmetry. The bookkeeping assignments are as follows (charges not listed are zero):

\begin{itemize}
\item Bifundamentals between $SU(N_{ci})$ and $SU(N_{cj})$: an oriented field $X_{ij}$ has $U(1)_{B_i}=+1$, $U(1)_{B_j}=-1$, while $X_{ji}$ has the opposite charges.
\item Quarks: $Q_i$ transforms as $(\mathbf{N}_{ci},\mathbf{N}_i)$ under $SU(N_{ci})\times SU(N_i)_L$ and has $U(1)_{B_i}=+1$. The field $\tilde Q_i$ transforms as $(\overline{\mathbf{N}}_{ci},\mathbf{N}_i)$ under $SU(N_{ci})\times SU(N_i)_R$ and has $U(1)_{B_i}=-1$.
\item Dual bifundamentals: $X_{ij}^*$ transforms as $(\overline{\mathbf{N}}_{ci}^{\rm d},\mathbf{N}_{cj}^{\rm d})$ and has $U(1)_{B_i}=+\frac{N_{ci}}{N_{ci}^{\rm d}}$, $U(1)_{B_j}=-\frac{N_{cj}}{N_{cj}^{\rm d}}$.
\item Dual quarks: $q_i$ transforms as $(\mathbf{N}_{ci}^{\rm d},\overline{\mathbf{N}}_i)$ under $SU(N_{ci}^{\rm d})\times SU(N_i)_R$ and has $U(1)_{B_i}=-\frac{N_{ci}}{N_{ci}^{\rm d}}$. The field $\tilde q_i$ transforms as $(\overline{\mathbf{N}}_{ci}^{\rm d},\overline{\mathbf{N}}_i)$ under $SU(N_{ci}^{\rm d})\times SU(N_i)_L$ and has $U(1)_{B_i}=+\frac{N_{ci}}{N_{ci}^{\rm d}}$.
\item Mesons and other gauge singlets: all $U(1)_{B_i}$ charges are zero.
\end{itemize}
The magnetic signs are chosen so that the baryon made from the left-flavor field $\tilde q_i$ has the same charge as the electric baryon made from $Q_i$. Using the normalization in which the Dynkin index of the fundamental representation is $T(\mathbf N_i)=\frac{1}{2}$, the flavor-baryon anomalies are
\begin{equation}\label{eq:flavor-baryon-anomalies}
\begin{aligned}
\Tr SU(N_i)_L^2B_i\big|_E
&=\Tr SU(N_i)_L^2B_i\big|_M=\frac{N_{ci}}{2},\\
\Tr SU(N_i)_R^2B_i\big|_E
&=\Tr SU(N_i)_R^2B_i\big|_M=-\frac{N_{ci}}{2}.
\end{aligned}
\end{equation}
In the quivers considered below, fields with opposite baryon charges occur in pairs, so the anomalies $\Tr B$, $\Tr B^3$, and $\Tr R^2B$ vanish in both the electric and magnetic theories.

\subsection{Cubic flavor anomalies}

With the convention stated above, a fundamental of a flavor $SU(N)$ contributes $+1$ to the cubic anomaly, while an antifundamental contributes $-1$. For $SU(N_i)_L$, the electric contribution comes from $Q_i$:
\begin{equation*}
\Tr SU(N_i)_L^3\Big|_E=N_{ci}.
\end{equation*}
In the magnetic theory, $\tilde q_i$ is an antifundamental of $SU(N_i)_L$ and contributes $-N_{ci}^{\rm d}$. The mesons $M_{I^{ji}}$ are fundamentals of $SU(N_i)_L$ and contribute $\sum_j a_{ji}N_j$. Therefore
\begin{equation}\label{eq:su3-anomaly-left}
\Tr SU(N_i)_L^3\Big|_M=-N_{ci}^{\rm d}+\sum_{j=1}^n a_{ji}N_j.
\end{equation}

Similarly, for $SU(N_i)_R$, the electric field $\tilde Q_i$ is a fundamental, so
\begin{equation*}
\Tr SU(N_i)_R^3\Big|_E=N_{ci}.
\end{equation*}
The magnetic contribution is $-N_{ci}^{\rm d}$ from $q_i$ and $+\sum_j a_{ij}N_j$ from the mesons $M_{I^{ij}}$. Hence
\begin{equation}\label{eq:su3-anomaly-right}
\Tr SU(N_i)_R^3\Big|_M=-N_{ci}^{\rm d}+\sum_{j=1}^n a_{ij}N_j.
\end{equation}
Using $N_{ci}^{\rm d}=\sum_jb_{ij}N_j-N_{ci}$, cubic flavor-anomaly matching is therefore equivalent to
\begin{equation}\label{eq:cubic-flavor-conditions}
\sum_j(a_{ji}-b_{ij})N_j=0,\qquad
\sum_j(a_{ij}-b_{ij})N_j=0.
\end{equation}
If the flavor ranks $N_j$ are arbitrary and independent, these conditions require
\begin{equation}\label{eq:a-equals-b}
a_{ij}=a_{ji}=b_{ij}.
\end{equation}
Thus finite-rank cubic flavor anomalies can force equality of the two meson multiplicities even when the large-$N$ index treats the $(i,j)$ and $(j,i)$ channels independently. This remains a statement about meson channels, not an assumption that the electric arrow sets $X_{ij}$ and $X_{ji}$ are identical.

\subsection{The $\Tr R$ and $\Tr R^3$ anomalies}

The $\Tr R$ anomaly in the electric theory receives contributions from gauginos ($R=1$), adjoint and bifundamental chiral multiplets, and quarks. The beta-function conditions~\eqref{eq:RQi} reduce the sum to
\begin{equation}\label{eq:TrR-electric}
\Tr R\big|_E=-\mathbf{N_c}^T A\,\mathbf{N_c}-\tr A.
\end{equation}

For later use, define the meson sums
\begin{equation}\label{eq:meson-R-moments}
S_{ij}^{(\ell)}
:=\sum_{I\in\mathcal I^{ij}}
\bigl(R_{Q_i}+R_{Q_j}+I-1\bigr)^\ell,
\qquad \ell=1,3.
\end{equation}
The magnetic linear anomaly is
\begin{equation}\label{eq:TrR-magnetic}
\Tr R\big|_M
=-\mathbf{N_c^{\rm d}}^T A\,\mathbf{N_c^{\rm d}}-\tr A
+\sum_{i,j}N_iN_jS_{ij}^{(1)}.
\end{equation}
The equality of the electric and magnetic charge sets, together with the pairing condition~\eqref{eq:pairing}, pairs the charges within each channel around $\Delta/2$. Consequently,
\[
\sum_{i,j}N_iN_j\sum_{I\in\mathcal I^{ij}}I
=\frac{\Delta}{2}\sum_{i,j}a_{ij}N_iN_j.
\]
With the cubic flavor condition $a=b$, equations~\eqref{eq:RQi}, \eqref{eq:dual-ranks}, and \eqref{eq:TrR-magnetic} then give $\Tr R|_M=\Tr R|_E$.

The cubic $R$ anomaly is a separate condition because it depends on the individual $R$-charges rather than only on the matrix $A$. For the electric theory,
\begin{align}
\Tr R^3\big|_E={}&
\sum_i(N_{ci}^2-1)
\left[1+\sum_{X_i}(R_{X_i}-1)^3\right]\nonumber\\
&+\sum_{i\neq j}\sum_{X_{ij}}
N_{ci}N_{cj}(R_{X_{ij}}-1)^3
+2\sum_iN_{ci}N_i(R_{Q_i}-1)^3.
\label{eq:TrR3-electric}
\end{align}
The sum over $i\neq j$ counts each oriented bifundamental once. The magnetic expression is
\begin{align}
\Tr R^3\big|_M={}&
\sum_i((N_{ci}^{\rm d})^2-1)
\left[1+\sum_{X_i}(R_{X_i}-1)^3\right]\nonumber\\
&+\sum_{i\neq j}\sum_{X_{ij}}
N_{ci}^{\rm d}N_{cj}^{\rm d}(R_{X_{ij}}-1)^3
+2\sum_iN_{ci}^{\rm d}N_i(R_{q_i}-1)^3
+\sum_{i,j}N_iN_jS_{ij}^{(3)}.
\label{eq:TrR3-magnetic}
\end{align}
Equality of~\eqref{eq:TrR3-electric} and~\eqref{eq:TrR3-magnetic} must be verified for each finite-rank candidate.

\subsection{Mixed $\Tr U(1)_{B_i}^2 R$ anomalies}

For the $i$-th baryon symmetry, the mixed anomaly $\Tr U(1)_{B_i}^2 R$ receives contributions from quarks and bifundamentals. In the electric theory:
\begin{equation}\label{eq:U1R-electric}
\begin{aligned}
\Tr B_i^2R\big|_E={}&2N_iN_{ci}(R_{Q_i}-1)\\
&+N_{ci}\sum_{j\neq i}N_{cj}
\left[\sum_{X_{ij}}(R_{X_{ij}}-1)
+\sum_{X_{ji}}(R_{X_{ji}}-1)\right]
=-2N_{ci}^2A_{ii}.
\end{aligned}
\end{equation}

The dual baryon-charge assignments give
\begin{equation}\label{eq:U1R-magnetic}
\begin{aligned}
\Tr B_i^2R\big|_M
&=\left(\frac{N_{ci}}{N_{ci}^{\rm d}}\right)^2
\left[
2N_iN_{ci}^{\rm d}(R_{q_i}-1)\right.\\
&\hspace{32mm}\left.
+N_{ci}^{\rm d}\sum_{j\neq i}N_{cj}^{\rm d}
\left(\sum_{X_{ij}}(R_{X_{ij}}-1)
+\sum_{X_{ji}}(R_{X_{ji}}-1)\right)
\right]\\
&=-2N_{ci}^2 A_{ii}.
\end{aligned}
\end{equation}
The prefactor is the square of the dual baryon charge. The last equality uses the corresponding beta function. Likewise, for $i\neq j$,
\begin{equation}\label{eq:mixed-baryon-R}
\Tr B_iB_jR\big|_E
=\Tr B_iB_jR\big|_M
=-2N_{ci}N_{cj}A_{ij}.
\end{equation}
These equations apply to any anomaly-free linear combination of the baryon charges.

\subsection{Mixed flavor-$R$ anomalies}

The electric mixed flavor anomalies are
\begin{equation}\label{eq:flavor-R-electric}
\Tr SU(N_i)_L^2R\big|_E
=\Tr SU(N_i)_R^2R\big|_E
=\frac{1}{2}N_{ci}(R_{Q_i}-1).
\end{equation}
The cubic flavor condition~\eqref{eq:a-equals-b}, together with the charge pairing within each meson channel, gives
\[
S_{ij}^{(1)}=S_{ji}^{(1)}
=b_{ij}\left(R_{Q_i}+R_{Q_j}-1+\frac{\Delta}{2}\right).
\]
The two magnetic anomalies are therefore equal:
\begin{equation}\label{eq:flavor-R-magnetic}
\Tr SU(N_i)_L^2R\big|_M
=\Tr SU(N_i)_R^2R\big|_M
=\frac{1}{2}N_{ci}^{\rm d}(R_{q_i}-1)
+\frac{1}{2}\sum_jN_jS_{ji}^{(1)}.
\end{equation}
Equations~\eqref{eq:RQi} and~\eqref{eq:b-matrix} imply
\[
\sum_jN_jb_{ij}(R_{Q_j}-1)
=-\frac{\Delta+2}{2}N_{ci}.
\]
Using this identity together with the quark pairing~\eqref{eq:quark-pairing} and the dual ranks~\eqref{eq:dual-ranks}, one obtains
\[
N_{ci}^{\rm d}(R_{q_i}-1)+\sum_jN_jS_{ji}^{(1)}
=N_{ci}(R_{Q_i}-1).
\]
Thus~\eqref{eq:flavor-R-magnetic} reduces to the electric result~\eqref{eq:flavor-R-electric}, and both mixed flavor-$R$ anomalies match.

\subsection{Baryon spectrum matching}

Beyond anomaly matching, a proposed duality must also identify the electric and magnetic baryon spectra. Assuming the cubic flavor condition $a=b$, the completeness and independence of the dressed-quark basis introduced below, and the stated charge pairing between each relevant electric and magnetic channel, the baryon map is constructed as follows. We illustrate the construction for the $SU(N_{c1})$ baryon sector. The other gauge nodes are analogous.

The electric $SU(N_{c1})$ baryons are constructed from the dressed quarks
\begin{equation*}
Q_{I^{1i}}=U_{I^{1i}}Q_i,\qquad i=1,\dots,n,
\end{equation*}
where $U_{I^{1i}}$ are the undressed mesons. Let $\{I^{1i}_\alpha\}_{\alpha=1}^{a_{1i}}$ denote the set of undressed-meson $R$-charges in the $(1,i)$ channel, and let $r_{i,\alpha}$ be the occupation number of the dressed quark $Q_{I^{1i}_\alpha}$. We write $\{r\}=\{r_{i,\alpha}\}$ for the full set of occupation numbers. The $SU(N_{c1})$ baryon specified by $\{r\}$ is
\begin{equation*}
B_{\{r\}}=\prod_{i=1}^n\prod_{\alpha=1}^{a_{1i}}\bigl(Q_{I^{1i}_\alpha}\bigr)^{r_{i,\alpha}}.
\end{equation*}
The occupation numbers satisfy
\begin{equation*}
0\leq r_{i,\alpha}\leq N_i,
\qquad
\sum_{i=1}^n\sum_{\alpha=1}^{a_{1i}}r_{i,\alpha}=N_{c1}.
\end{equation*}
The first condition bounds the occupation of each dressed-quark species by the dimension of its flavor representation, while the second ensures that the baryon contains exactly $N_{c1}$ dressed quarks. Their gauge indices are contracted with the $SU(N_{c1})$ epsilon tensor.
The total number of such baryons is
\begin{equation*}
\binom{\sum_i a_{1i}N_i}{N_{c1}}.
\end{equation*}

The magnetic baryon $\tilde B_{\{\tilde r\}}$ dual to $B_{\{r\}}$ is built from the corresponding dressed $\tilde q$ fields. The undressed mesons in the dual theory are paired with those of the electric theory by $I^{1i}_\alpha+I^{1i\,\prime}_\alpha=\Delta$. After ordering the electric undressed-meson charges in channel $(1,i)$ as $I^{1i}_1,\dots,I^{1i}_{a_{1i}}$, we order the paired magnetic charges as
\begin{equation*}
I^{1i\,\prime}_\alpha=\Delta-I^{1i}_{a_{1i}+1-\alpha},\qquad \alpha=1,\dots,a_{1i}.
\end{equation*}
The dressed magnetic quarks are therefore
\begin{equation*}
\tilde q_{I^{1i\,\prime}_\alpha}=U^*_{I^{1i\,\prime}_\alpha}\tilde q_i,
\qquad
R(\tilde q_{I^{1i\,\prime}_\alpha})=R_{q_i}+I^{1i\,\prime}_\alpha.
\end{equation*}
For each flavor group $SU(N_i)_L$ and each meson species, the magnetic baryon uses the complementary set of flavor indices to the electric baryon. Equivalently, its occupation numbers are
\begin{equation*}
\tilde r_{i,\alpha}=N_i-r_{i,a_{1i}+1-\alpha},\qquad \alpha=1,\dots,a_{1i},
\end{equation*}
so that
\begin{equation*}
\tilde B_{\{\tilde r\}}=
\prod_{i=1}^n\prod_{\alpha=1}^{a_{1i}}
\bigl(\tilde q_{I^{1i\,\prime}_\alpha}\bigr)^{\tilde r_{i,\alpha}}.
\end{equation*}
Thus the dual baryon uses the complementary occupation numbers in the reversed order dictated by the meson pairing. Summing the dual occupation numbers gives
\begin{equation*}
\sum_{i=1}^n\sum_{\alpha=1}^{a_{1i}}\tilde r_{i,\alpha}
=\sum_i a_{1i}N_i-\sum_{i,\alpha}r_{i,\alpha}
=\sum_i a_{1i}N_i-N_{c1}=N_{c1}^{\rm d},
\end{equation*}
confirming that the magnetic operator contains exactly $N_{c1}^{\rm d}$ dressed quarks. The total number of magnetic baryons is
\begin{equation*}
\binom{\sum_i a_{1i}N_i}{N_{c1}^{\rm d}}=\binom{\sum_i a_{1i}N_i}{\sum_i a_{1i}N_i-N_{c1}}=\binom{\sum_i a_{1i}N_i}{N_{c1}},
\end{equation*}
identical to the electric count.

The baryon charges also agree. The charges of the bifundamentals cancel successively along a dressed path, so each electric dressed quark ending at node~1 has $B_1=+1$, whereas each corresponding magnetic dressed $\tilde q$ has $B_1=N_{c1}/N_{c1}^{\rm d}$. Therefore
\begin{equation}\label{eq:baryon-charge-matching}
B_1\bigl(B_{\{r\}}\bigr)=N_{c1}
=B_1\bigl(\tilde B_{\{\tilde r\}}\bigr).
\end{equation}

Furthermore, the $R$-charges of the baryons also match. The $R$-charge of an electric baryon is
\begin{equation*}
R(B_{\{r\}})=\sum_{i=1}^n\sum_{\alpha=1}^{a_{1i}}r_{i,\alpha}\bigl(R_{Q_i}+I^{1i}_\alpha\bigr).
\end{equation*}
The magnetic $R$-charge is
\begin{align*}
R(\tilde B_{\{\tilde r\}})
&=\sum_{i=1}^n\sum_{\alpha=1}^{a_{1i}}
\tilde r_{i,\alpha}\bigl(R_{q_i}+I^{1i\,\prime}_\alpha\bigr)\\
&=\sum_{i=1}^n\sum_{\alpha=1}^{a_{1i}}
\bigl(N_i-r_{i,\alpha}\bigr)
\bigl(R_{q_i}+\Delta-I^{1i}_\alpha\bigr),
\end{align*}
where the second line is obtained by relabeling $\alpha\to a_{1i}+1-\alpha$. Since the magnetic gauge sector has the same undressed-meson spectrum in the paired channel, the two lists of $R$-charges $\{I^{1i}_\alpha\}$ and $\{I^{1i\,\prime}_\alpha\}$ have the same total $R$-charge. Combining this with $I^{1i}_\alpha+I^{1i\,\prime}_{a_{1i}+1-\alpha}=\Delta$ gives
\begin{equation*}
\sum_{\alpha=1}^{a_{1i}}I^{1i}_\alpha=\frac{\Delta}{2}a_{1i}.
\end{equation*}
Using this identity, the quark pairing condition~\eqref{eq:quark-pairing}, the electric beta-function relation~\eqref{eq:RQi}, and $a_{ij}=b_{ij}=\frac{2+\Delta}{2}(A^{-1})_{ij}$, the difference between the magnetic and electric baryon $R$-charges becomes
\begin{align*}
R(\tilde B_{\{\tilde r\}})-R(B_{\{r\}})
&=\sum_i a_{1i}N_i\Bigl(R_{q_i}+\frac{\Delta}{2}\Bigr)
-\sum_{i,\alpha}r_{i,\alpha}\bigl(R_{Q_i}+R_{q_i}+\Delta\bigr)\\
&=\sum_i a_{1i}N_i(1-R_{Q_i})
-\Bigl(1+\frac{\Delta}{2}\Bigr)N_{c1}\\
&=\sum_{i,j}a_{1i}A_{ij}N_{cj}
-\Bigl(1+\frac{\Delta}{2}\Bigr)N_{c1}\\
&=0.
\end{align*}
Under the assumptions stated above, this gives a one-to-one correspondence between the electric and magnetic baryons: the two spectra contain the same number of operators, and the electric and magnetic baryons in each dual pair have the same baryon charge and $R$-charge.

\section{Examples of quiver dualities}
\label{sec:examples}

In this section, we illustrate the framework with two known two-node families containing one adjoint chiral at each gauge node. The first is the symmetric $SU$--$SU$ family of Brodie and Brodie--Hanany. The second is the $SO$--$USp$ product-group family of Ahn, Oh, and Tatar. The $SU$ example displays the complex matrix-index derivation in detail, while the $SO$--$USp$ example applies the real-Gaussian kernel and finite-rank anomaly conditions.

\subsection{Two-node $SU$--$SU$ duality}
\label{sec:two-node-SU}

\subsubsection{Setup}

We consider the general symmetric two-node quiver shown in Figure~\ref{fig:two-node-general}. The electric theory has gauge group $SU(N_{c1})\times SU(N_{c2})$, one adjoint chiral $X_1$ of $SU(N_{c1})$, one adjoint chiral $X_2$ of $SU(N_{c2})$, and a bifundamental pair
\[
X:(\mathbf{N}_{c1},\overline{\mathbf{N}}_{c2}),\qquad
\tilde X:(\overline{\mathbf{N}}_{c1},\mathbf{N}_{c2}).
\]
The flavor fields are, with flavor ranks denoted by $N_1$ and $N_2$,
\[
\begin{array}{ll}
Q_1:(\mathbf{N}_{c1},\mathbf{N}_1)\ \text{under }SU(N_{c1})\times SU(N_1)_L,
& \tilde Q_1:(\overline{\mathbf{N}}_{c1},\mathbf{N}_1)\ \text{under }SU(N_{c1})\times SU(N_1)_R,\\[2pt]
Q_2:(\mathbf{N}_{c2},\mathbf{N}_2)\ \text{under }SU(N_{c2})\times SU(N_2)_L,
& \tilde Q_2:(\overline{\mathbf{N}}_{c2},\mathbf{N}_2)\ \text{under }SU(N_{c2})\times SU(N_2)_R.
\end{array}
\]
The magnetic theory contains the dual gauge groups $SU(N_{c1}^{\rm d})\times SU(N_{c2}^{\rm d})$, dual adjoints $X_1^*,X_2^*$, the dual bifundamental pair $X^*,\tilde X^*$, the dual quarks $q_i,\tilde q_i$, and elementary singlets
\[
M_{I^{11}},\quad M_{I^{22}},\quad M_{I^{12}},\quad M_{I^{21}}.
\]
The superpotential is
\begin{equation}\label{eq:W-two-node}
W=\Tr X_1^{p+1}+(-1)^{p+1}\Tr X_2^{p+1}
+\Tr(X_1 X\tilde X)+\Tr(X_2\tilde X X),
\end{equation}
where $p\ge1$. The $R$-charges are fixed by this superpotential:
\begin{equation}\label{eq:R-two-node}
R_{X_1}=R_{X_2}=r=\frac{2}{p+1},\qquad
R_X=R_{\tilde X}=s=1-\frac{r}{2}=\frac{p}{p+1}.
\end{equation}
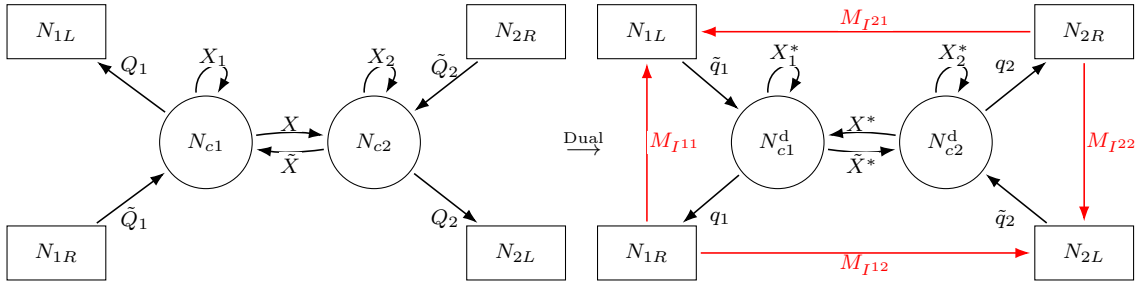
\begin{figure}[H]
    \centering
    \begingroup
    \large
    \resizebox{\textwidth}{!}{
    \begin{tikzpicture}[
        gauge/.style={draw,circle,minimum size=3.0em,inner sep=1pt,align=center},
        flavor/.style={draw,rectangle,minimum width=3.2em,minimum height=1.75em,inner sep=2pt,align=center},
        field/.style={semithick,-{Latex[length=2.0mm,width=1.4mm]},shorten >=1.5pt,shorten <=1.5pt},
        meson/.style={red,semithick,-{Latex[length=2.0mm,width=1.4mm]},shorten >=1.5pt,shorten <=1.5pt},
        flab/.style={fill=white,inner sep=0.5pt},
        qlab/.style={inner sep=0.25pt},
        every node/.style={font=\scriptsize}
    ]
        \node[flavor] (e1L) at (-2.00,3.28) {$N_{1L}$};
        \node[flavor] (e1R) at (-2.00,0.32) {$N_{1R}$};
        \node[gauge] (e1) at (0,1.80) {$N_{c1}$};
        \node[gauge] (e2) at (2.25,1.80) {$N_{c2}$};
        \node[flavor] (e2R) at (4.15,3.28) {$N_{2R}$};
        \node[flavor] (e2L) at (4.15,0.32) {$N_{2L}$};

        \draw[field] (e1.102) .. controls (-0.15,2.92) and (0.42,2.92) .. node[above,flab] {$X_1$} (e1.78);
        \draw[field] (e2.102) .. controls (2.10,2.92) and (2.67,2.92) .. node[above,flab] {$X_2$} (e2.78);
        \draw[field] (e1) -- (e1L);
        \node[qlab] at (-0.95,2.82) {$Q_1$};
        \draw[field] (e1R) -- (e1);
        \node[qlab] at (-0.95,0.78) {$\tilde Q_1$};
        \draw[field,bend left=8] (e1) to node[above,flab] {$X$} (e2);
        \draw[field,bend left=8] (e2) to node[below,flab] {$\tilde X$} (e1);
        \draw[field] (e2R) -- (e2);
        \node[qlab] at (3.20,2.82) {$\tilde Q_2$};
        \draw[field] (e2) -- (e2L);
        \node[qlab] at (3.20,0.78) {$Q_2$};

        \node at (5.05,1.80) {$\stackrel{\text{Dual}}{\longrightarrow}$};

        \node[flavor] (m1L) at (5.90,3.28) {$N_{1L}$};
        \node[flavor] (m1R) at (5.90,0.32) {$N_{1R}$};
        \node[gauge] (m1) at (7.65,1.80) {$N_{c1}^{\rm d}$};
        \node[gauge] (m2) at (9.90,1.80) {$N_{c2}^{\rm d}$};
        \node[flavor] (m2R) at (11.75,3.28) {$N_{2R}$};
        \node[flavor] (m2L) at (11.75,0.32) {$N_{2L}$};

        \draw[field] (m1.102) .. controls (7.50,2.92) and (8.07,2.92) .. node[above,flab] {$X_1^*$} (m1.78);
        \draw[field] (m2.102) .. controls (9.75,2.92) and (10.32,2.92) .. node[above,flab] {$X_2^*$} (m2.78);
        \draw[field] (m1L) -- (m1);
        \node[qlab] at (6.88,2.82) {$\tilde q_1$};
        \draw[field] (m1) -- (m1R);
        \node[qlab] at (6.88,0.78) {$q_1$};
        \draw[field,bend right=8] (m2) to node[above,flab] {$X^*$} (m1);
        \draw[field,bend right=8] (m1) to node[below,flab] {$\tilde X^*$} (m2);
        \draw[field] (m2) -- (m2R);
        \node[qlab] at (10.70,2.82) {$q_2$};
        \draw[field] (m2L) -- (m2);
        \node[qlab] at (10.70,0.78) {$\tilde q_2$};

        \draw[meson] (m1R) -- node[right,flab] {$M_{I^{11}}$} (m1L);
        \draw[meson] (m2R) -- node[right,flab] {$M_{I^{22}}$} (m2L);
        \draw[meson] (m1R) -- node[below,flab] {$M_{I^{12}}$} (m2L);
        \draw[meson] (m2R) -- node[above,flab] {$M_{I^{21}}$} (m1L);
    \end{tikzpicture}}
    \endgroup
    \caption{Seiberg duality for the general symmetric two-node $SU(N_{c1})\times SU(N_{c2})$ quiver. The electric theory contains adjoints $X_1,X_2$, a bifundamental pair $X,\tilde X$, and flavor groups $SU(N_i)_{L,R}$. The magnetic theory contains the corresponding dual fields. The red arrows in the magnetic quiver denote elementary gauge-singlet mesons $M_{I^{11}},M_{I^{22}},M_{I^{12}},M_{I^{21}}$, whose $R$-charge sets follow from~\eqref{eq:meson-general}.}
    \label{fig:two-node-general}
\end{figure}

\subsubsection{Matrix Hilbert series and the constraint equation}

The matrix $M_Q(t)$ encoding the gauge-sector $R$-charges is
\begin{equation*}
M_Q(t)=\begin{pmatrix}
t^{r} & t^{s}\\[2pt]
t^{s} & t^{r}
\end{pmatrix},
\qquad r=\frac{2}{p+1},\quad s=\frac{p}{p+1}.
\end{equation*}
Introduce
\[
y=t^{1/(p+1)}
\]
so that all powers become integral powers of $y$. By the definition~\eqref{eq:Hilbert-series-def},
\begin{align*}
H(Q,t=y^{p+1})^{-1}
&=\left(1-t^2-M_Q(t)+t^2M_Q^T(t^{-1})\right)\Big|_{t=y^{p+1}}\\
&=\begin{pmatrix}
(1-y^2)(1+y^{2p}) & -y^p(1-y^2)\\
-y^p(1-y^2) & (1-y^2)(1+y^{2p})
\end{pmatrix}.
\end{align*}
A direct two-by-two inversion gives the reduced entries of \(H(Q,t=y^{p+1})\):
\begin{align*}
H_{11}=H_{22}&=\frac{1+y^{2p}}{(1-y^2)(1+y^{2p}+y^{4p})},&
H_{12}=H_{21}&=\frac{y^p}{(1-y^2)(1+y^{2p}+y^{4p})}.
\end{align*}
Here $H(Q,t)$ is symmetric. The left-hand side of~\eqref{eq:dual-matrix-form} therefore becomes $(1-y^L)H(Q,t=y^{p+1})$. All entries have the same reduced denominator
\[
q(y)=(1-y^2)(1+y^{2p}+y^{4p})
=\Phi_1(y)\Phi_2(y)\prod_{\substack{d\text{ divides }6p\\ d\text{ does not divide }2p}}\Phi_d(y).
\]
Every factor in this reduced denominator appears only once, and the numerators in the reduced entries have no common factors with $q(y)$. Thus the polynomial condition says that every factor in $q(y)$ must be canceled by $1-y^L$. This holds precisely when $L$ is divisible by all the cyclotomic orders appearing in $q(y)$. The relevant orders and the smallest allowed value of $L$ are
\begin{equation}\label{eq:Delta-general}
\begin{gathered}
\mathcal{N}=\{1,2\}\cup\{d:d\text{ divides }6p\text{ and }d\text{ does not divide }2p\},\\
L=\operatorname{lcm}(\mathcal{N}),\qquad \Delta+2=\frac{L}{p+1}.
\end{gathered}
\end{equation}
Here $\operatorname{lcm}$ means the smallest positive integer divisible by every number in $\mathcal N$. Since $6p\in\mathcal N$ and every element of $\mathcal N$ divides $6p$, equation~\eqref{eq:Delta-general} gives $L=6p$.

Substituting $L=6p$ into the matrix constraint and simplifying gives the meson polynomials directly:
\begin{equation}\label{eq:meson-general}
\boxed{
\begin{aligned}
(1-y^{6p})H_{11}(y^{p+1})&=
\frac{(1-y^{6p})(1+y^{2p})}{(1-y^2)(1+y^{2p}+y^{4p})}
=\sum_{\ell=0}^{2p-1}y^{2\ell},\\
(1-y^{6p})H_{12}(y^{p+1})&=
\frac{(1-y^{6p})y^p}{(1-y^2)(1+y^{2p}+y^{4p})}
=\sum_{\ell=0}^{p-1}y^{p+2\ell}.
\end{aligned}}
\end{equation}
The final sums are finite polynomials with nonnegative integer coefficients. The diagonal polynomial contains $2p$ monomials, while the off-diagonal polynomial contains $p$ monomials. Since a monomial $y^a$ corresponds to the $R$-charge $I^{ij}=a/(p+1)$, writing $a_{ij}=|\mathcal I^{ij}|$ gives the general result
\begin{equation}\label{eq:su-general-result}
\boxed{
L=6p,\qquad
\Delta+2=\frac{6p}{p+1},\qquad
a_{11}=a_{22}=2p,\qquad
a_{12}=a_{21}=p.}
\end{equation}

For $p=1$, equation~\eqref{eq:meson-general} gives
\[
\mathcal I^{11}=\mathcal I^{22}=\{0,1\},
\qquad
\mathcal I^{12}=\mathcal I^{21}=\left\{\frac{1}{2}\right\}.
\]
In this case $r=1$ and $s=\frac{1}{2}$. The adjoints are massive and can be integrated out, recovering the Brodie--Hanany duality~\cite{Brodie:1997sz}. For general $p$, the meson multiplicities and magnetic gauge ranks reproduce the product-group duality with adjoint matter found by Brodie~\cite{Brodie:1996vx}. Thus the large-$N$ index constraint recovers this known $SU$--$SU$ duality family.

\subsubsection{Dual gauge group and meson multiplicities}

With $N_1$ and $N_2$ fundamental flavors, the NSVZ beta functions~\eqref{eq:beta-matrix} determine the quark $R$-charges and, via the pairing condition $2R_{Q_i}+2R_{q_i}+\Delta=2$, the dual gauge ranks. The $2\times2$ matrix $A$ from~\eqref{eq:A-matrix} is
\begin{equation*}
A=\begin{pmatrix}
r & s-1\\
s-1 & r
\end{pmatrix}
=\frac{1}{p+1}\begin{pmatrix}
2 & -1\\
-1 & 2
\end{pmatrix},
\qquad
A^{-1}=\frac{p+1}{3}\begin{pmatrix}
2 & 1\\
1 & 2
\end{pmatrix}.
\end{equation*}
The beta-function coefficients~\eqref{eq:b-matrix} are
\begin{equation*}
b_{11}=b_{22}=\frac{(\Delta+2)(p+1)}{3}=2p,\qquad
b_{12}=b_{21}=\frac{(\Delta+2)(p+1)}{6}=p.
\end{equation*}
The sums of coefficients in~\eqref{eq:meson-general} give the same values, so the cubic flavor condition~\eqref{eq:a-equals-b} is satisfied and $a_{ij}=b_{ij}$ in this family. The dual gauge ranks then follow from~\eqref{eq:dual-ranks}:
\begin{equation}\label{eq:dual-ranks-two-node}
\boxed{\begin{aligned}
N_{c1}^{\rm d}&=a_{11}N_1+a_{12}N_2-N_{c1}
=2pN_1+pN_2-N_{c1},\\
N_{c2}^{\rm d}&=a_{12}N_1+a_{11}N_2-N_{c2}
=pN_1+2pN_2-N_{c2}.
\end{aligned}}
\end{equation}
For the first few values:
\begin{center}
\begin{tabular}{c|c|c|c}
$p$ & $\Delta+2$ & $a_{11}=a_{22}$ & $a_{12}=a_{21}$ \\
\hline
$1$ & $3$ & $2$ & $1$ \\
$2$ & $4$ & $4$ & $2$ \\
$3$ & $9/2$ & $6$ & $3$ \\
\end{tabular}
\end{center}
These multiplicities have a direct path interpretation. The coefficient $a_{11}$ counts the independent undressed meson paths that begin and end at node~1, constructed from $X_1$ and the closed path $X\tilde X$, whereas $a_{12}$ counts the independent undressed paths from node~2 to node~1. For $p=1$, the diagonal paths are $1$ and $X\tilde X$, while the only off-diagonal path is $X$. Hence $a_{11}=2$ and $a_{12}=1$.

\subsubsection{Truncation of the chiral ring by the superpotential}

The F-term equations of the superpotential~\eqref{eq:W-two-node} provide a chiral-ring consistency check of the finite spectra in~\eqref{eq:meson-general}:
\begin{equation}\label{eq:F-term-two}
X_1^{p}+X\tilde X=0,\qquad
(-1)^{p+1}X_2^{p}+\tilde X X=0,\qquad
\tilde X X_1+X_2\tilde X=0,\qquad
X_1 X+X X_2=0.
\end{equation}
Let $Y=X\tilde X$. The first equation gives $Y=-X_1^p$, while the second gives $\tilde X X=(-1)^pX_2^p$. Moving $X_2^p$ through $X$ with $X_1X+XX_2=0$ then yields
\[
Y^2=X(\tilde X X)\tilde X
=(-1)^pXX_2^p\tilde X
=X_1^pX\tilde X=-Y^2,
\]
and hence $Y^2=0$. Therefore $X_1^{2p}=0$ in the chiral ring. A basis of undressed diagonal paths starting and ending on node~1 is
\begin{equation*}
1,\;X_1,\;X_1^2,\;\dots,\;X_1^{2p-1},
\end{equation*}
with $R$-charges $\{0,\frac{2}{p+1},\frac{4}{p+1},\dots,\frac{2(2p-1)}{p+1}\}$. This gives $2p$ diagonal undressed mesons, in agreement with the diagonal polynomial obtained from the first line of~\eqref{eq:meson-general}.

Similarly, the off-diagonal paths occur in two orientations. Paths from node~2 to node~1 may be represented by $X_1^\ell X$, while paths from node~1 to node~2 may be represented by $X_2^\ell\tilde X$, with $\ell=0,\dots,p-1$. The last two relations in~\eqref{eq:F-term-two} allow adjoint powers to be moved between the two ends of each bifundamental path. Both orientations have $R$-charges
\[
s+\ell r=\frac{p+2\ell}{p+1},
\]
which are exactly the monomials in the off-diagonal entries of~\eqref{eq:meson-general}. For $p=2$, these charges are $\{\frac23,\frac43\}$, while for $p=3$ they are $\{\frac34,\frac54,\frac74\}$.

Together, the diagonal and off-diagonal path bases reproduce the finite meson spectra predicted by the constraint equation~\eqref{eq:dual-matrix-form}. The relative sign in~\eqref{eq:W-two-node} is essential for the classical F-term truncation~\eqref{eq:F-term-two} to hold for both odd and even $p$, as in the product-group duality of Brodie~\cite{Brodie:1996vx}. This agreement is a chiral-ring consistency check rather than an independent proof of duality. The finite-rank anomaly and baryon conditions of Section~\ref{sec:anomalies} are also satisfied for this family.

\subsubsection{Finite-rank SCI check}

As an additional finite-rank check, set all flavor fugacities to \(1\), so the index depends only on \(t\) and the spacetime fugacity \(x\). For \(p=1\), the small-\(t\) expansion is evaluated at
\[
N_{c1}=N_{c2}=4,\qquad N_1=N_2=3.
\]
The dual ranks and beta-function constraints give \(N_{c1}^{\rm d}=N_{c2}^{\rm d}=5\), \(R_{Q_1}=R_{Q_2}=\frac{1}{3}\), and \(R_{q_1}=R_{q_2}=\frac{1}{6}\). The finite-rank electric and magnetic \(SU\) indices agree:
\begin{equation}\label{eq:su-sci-expansion}
\mathcal I_E^{SU}=\mathcal I_M^{SU}
=1+18t^{\frac{2}{3}}+t+18t^{\frac{7}{6}}+171t^{\frac{4}{3}}
+\frac{18(1+x)^2}{x}t^{\frac{5}{3}}+\cdots .
\end{equation}
The agreement of the SCI expansions provides a nontrivial finite-\(N\) check of this recovered \(SU\)--\(SU\) duality.

\subsection{Two-node $SO$--$USp$ duality}
\label{sec:sosp-example}

The $SO$--$USp$ counterpart of the symmetric two-node family in Section~\ref{sec:two-node-SU} has one adjoint chiral at each gauge node. This product-group duality was discussed by Ahn, Oh, and Tatar~\cite{Ahn:1997gs}. We derive its large-$N$ kernel and magnetic ranks, and then describe its electric superpotential, meson spectrum, and chiral-ring truncation.

\subsubsection{Electric setup}

Let $p=2k+1$ be a positive odd integer and define
\begin{equation}\label{eq:sosp-two-adjoint-rs}
r=\frac{2}{p+1},\qquad s=\frac{p}{p+1}.
\end{equation}
The restriction to odd $p$ ensures that $p+1=2(k+1)$ is even, as required for a nonvanishing trace power of the antisymmetric $SO$ adjoint. The electric gauge group is
\begin{equation*}
SO(N_{c1})\times USp(N_{c2}),
\end{equation*}
where the fundamental dimension $N_{c2}$ of the $USp$ factor is even. The flavor symmetry is $SU(N_1)\times SU(N_2)$. The electric matter consists of
\begin{itemize}
\item an antisymmetric adjoint $X_1$ of $SO(N_{c1})$, with $R_{X_1}=r$,
\item a symmetric adjoint $X_2$ of $USp(N_{c2})$, with $R_{X_2}=r$,
\item a bifundamental $X$ in $(\mathbf N_{c1},\mathbf N_{c2})$, with $R_X=s$,
\item quarks $Q_1$ and $Q_2$ in $(\mathbf N_{c1},\mathbf N_1)$ and $(\mathbf N_{c2},\mathbf N_2)$, respectively.
\end{itemize}

\subsubsection{Large-$N$ constraint and dual gauge group}

The denominator-cleared quadratic kernel defined in Appendix~\ref{app:sosp} is
\begin{equation}\label{eq:sosp-two-adjoint-kernel}
\mathcal K_p(t)=
\begin{pmatrix}
1-t^2-\bigl(t^r-t^{2-r}\bigr)&-\bigl(t^s-t^{2-s}\bigr)\\[3pt]
-\bigl(t^s-t^{2-s}\bigr)&1-t^2-\bigl(t^r-t^{2-r}\bigr)
\end{pmatrix}.
\end{equation}
The antisymmetric $SO$ adjoint and the symmetric $USp$ adjoint give the same diagonal quadratic term. Their different flavor parities are encoded in the $O(N)$ even-mode terms of the index.

As in the $SU$ example, set $y=t^{1/(p+1)}$. Equation~\eqref{eq:sosp-two-adjoint-kernel} becomes
\begin{equation*}
\mathcal K_p(y^{p+1})=(1-y^2)
\begin{pmatrix}
1+y^{2p}&-y^p\\
-y^p&1+y^{2p}
\end{pmatrix}.
\end{equation*}
The $SO/USp$ matrix constraint~\eqref{eq:sosp-meson-matrix-constraint} is satisfied for $\Delta+2=6p/(p+1)$. Indeed,
\begin{equation}\label{eq:sosp-two-adjoint-index-constraint}
\boxed{
(1-y^{6p})\mathcal K_p(y^{p+1})^{-1}
=\begin{pmatrix}
\displaystyle\sum_{j=0}^{2p-1}y^{2j}&
\displaystyle\sum_{\ell=0}^{p-1}y^{p+2\ell}\\[10pt]
\displaystyle\sum_{\ell=0}^{p-1}y^{p+2\ell}&
\displaystyle\sum_{j=0}^{2p-1}y^{2j}
\end{pmatrix}.}
\end{equation}
Thus
\begin{equation}\label{eq:sosp-two-adjoint-index-data}
\Delta+2=\frac{6p}{p+1},\qquad
a_{11}=a_{22}=2p,\qquad a_{12}=a_{21}=p.
\end{equation}
The diagonal entries give the paths $1,X_i,\ldots,X_i^{2p-1}$, while the off-diagonal entries give $X_1^\ell X\simeq XX_2^\ell$ for $\ell=0,\ldots,p-1$.

The $O(N)$ terms distinguish the symmetric and antisymmetric flavor representations of the diagonal mesons. The gauge-index contractions give
\begin{equation*}
\begin{array}{c|cc}
\text{gauge node}&j\ \text{even}&j\ \text{odd}\\
\hline
SO&\text{flavor-symmetric}&\text{flavor-antisymmetric}\\
USp&\text{flavor-antisymmetric}&\text{flavor-symmetric}.
\end{array}
\end{equation*}
The $SO$ rule follows from $(X_1^j)^T=(-1)^jX_1^j$. At a $USp$ node, contraction with the antisymmetric symplectic invariant contributes one additional minus sign and reverses the rule.
For example, the electric meson $Q_1X_1^jQ_1$ becomes an elementary magnetic singlet and couples through~\eqref{eq:deltaw} to a magnetic composite $q_1(X_1^*)^{j'}q_1$, where
\begin{equation*}
j'=2p-1-j.
\end{equation*}
Since $2p-1$ is odd, $j'$ has the opposite parity from $j$. The singlet and the magnetic composite must have the same flavor-symmetry type for their contraction in the superpotential to be nonzero. If $q_1$ were attached to another $SO$ node, the opposite parity of $j'$ would give the wrong symmetry type. Attaching $q_1$ to a $USp$ node reverses the symmetry rule in the table, so the two reversals cancel and the flavor representations match. Applying the same argument to the electric $USp$ node places $q_2$ on an $SO$ node. We therefore order the magnetic nodes by their flavor labels and write
\begin{equation}\label{eq:sosp-two-adjoint-dual-group}
USp(N_{c1}^{\rm d})\times SO(N_{c2}^{\rm d}).
\end{equation}
Thus, with the flavor labels held fixed, the duality exchanges the gauge-group types $SO\leftrightarrow USp$ at the two nodes. If both products are instead ordered as $SO\times USp$, the two nodes and their attached flavor groups are exchanged.

The finite ranks follow from the cubic flavor anomalies. Each diagonal tower contains $p$ symmetric and $p$ antisymmetric representations. Using $A_{SU(N)}(\operatorname{Sym}^2)=N+4$ and $A_{SU(N)}(\wedge^2)=N-4$, anomaly matching gives
\begin{align*}
N_{c1}&=-N_{c1}^{\rm d}+p(N_1+4)+p(N_1-4)+pN_2,\\
N_{c2}&=-N_{c2}^{\rm d}+p(N_2+4)+p(N_2-4)+pN_1.
\end{align*}
Hence
\begin{equation}\label{eq:sosp-two-adjoint-dual-ranks}
\boxed{
N_{c1}^{\rm d}=2pN_1+pN_2-N_{c1},\qquad
N_{c2}^{\rm d}=pN_1+2pN_2-N_{c2}.}
\end{equation}
This is the same rank formula as in the $SU$--$SU$ example, equation~\eqref{eq:dual-ranks-two-node}, although the magnetic gauge-group types are different.\footnote{In the conventions of~\cite{Ahn:1997gs}, where the fundamental of $Sp(n)$ has dimension $2n$ and $N_i=2N_{fi}$, equation~\eqref{eq:sosp-two-adjoint-dual-ranks} is their magnetic-rank formula.}
Both ranks in~\eqref{eq:sosp-two-adjoint-dual-ranks} must be nonnegative. The quantity $N_{c1}^{\rm d}$ is the fundamental dimension of the magnetic $USp$ factor and must be even. The electric $USp$ global-anomaly condition requires $N_2+N_{c1}$ to be even~\cite{Witten:1982fp}, which ensures this parity because $p$ is odd. The magnetic $USp$ node has $N_1+N_{c2}^{\rm d}$ fundamental chirals, an even number because $p+1$ and $N_{c2}$ are even.

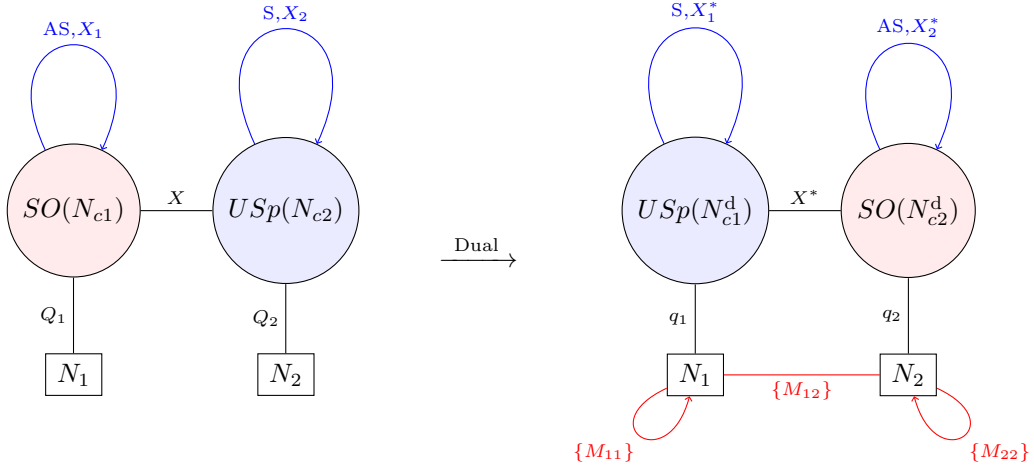
\begin{figure}[H]
\centering\small
\begin{center}
\begin{tikzcd}[column sep=2.7em,row sep=2.6em]
|[draw,circle,fill=red!8]| SO(N_{c1})
\ar[blue,loop,looseness=7,out=115,in=65,"{\textcolor{blue}{\mathrm{AS}},X_1}"]
\ar[r,"X",no head]
&|[draw,circle,fill=blue!8]| USp(N_{c2})
\ar[blue,loop,looseness=7,out=115,in=65,"{\textcolor{blue}{\mathrm{S}},X_2}"]\\
|[draw,rectangle]|N_1\ar[u,"Q_1",no head]
&|[draw,rectangle]|N_2\ar[u,"Q_2",no head]
\end{tikzcd}
\hspace{1.2em}
$\xrightarrow{\ \text{Dual}\ }$
\hspace{1.2em}
\begin{tikzcd}[column sep=2.7em,row sep=2.6em]
|[draw,circle,fill=blue!8]| USp(N_{c1}^{\rm d})
\ar[blue,loop,looseness=7,out=115,in=65,"{\textcolor{blue}{\mathrm{S}},X_1^*}"]
\ar[r,"X^*",no head]
&|[draw,circle,fill=red!8]| SO(N_{c2}^{\rm d})
\ar[blue,loop,looseness=7,out=115,in=65,"{\textcolor{blue}{\mathrm{AS}},X_2^*}"]\\
|[draw,rectangle]|N_1\ar[u,"q_1",no head]
\ar[red,loop,looseness=9,out=205,in=255,"{\color{red}\{M_{11}\}}"']
\ar[r,red,"{\{M_{12}\}}"',no head]
&|[draw,rectangle]|N_2\ar[u,"q_2",no head]
\ar[red,loop,looseness=9,out=335,in=285,"{\color{red}\{M_{22}\}}"]
\end{tikzcd}
\end{center}
\caption{The two-adjoint $SO$--$USp$ product-group duality~\cite{Ahn:1997gs}. Light red and light blue nodes denote $SO$ and $USp$ gauge groups. With the flavor labels held fixed, the gauge-group types are exchanged in the magnetic theory. The blue loops denote antisymmetric $SO$ adjoints and symmetric $USp$ adjoints. The red lines denote the three sets of elementary mesons. Matter lines have no arrowheads because the gauge representations are self-conjugate.}
\label{fig:sosp-example}
\end{figure}

The magnetic charged fields are a symmetric adjoint $X_1^*$ of $USp(N_{c1}^{\rm d})$, an antisymmetric adjoint $X_2^*$ of $SO(N_{c2}^{\rm d})$, and a bifundamental $X^*$. The quarks $q_1$ and $q_2$ are fundamentals of $USp(N_{c1}^{\rm d})$ and $SO(N_{c2}^{\rm d})$, respectively, and antifundamentals of $SU(N_1)$ and $SU(N_2)$. The dual quivers are shown in Figure~\ref{fig:sosp-example}.

\subsubsection{Superpotential, meson spectrum, and chiral-ring truncation}

Let $J_2$ be the invariant antisymmetric tensor of $USp(N_{c2})$ and define
\begin{equation*}
\widetilde X=J_2X^T.
\end{equation*}
The electric superpotential is~\cite{Ahn:1997gs}
\begin{equation}\label{eq:sosp-two-adjoint-electric-W}
\begin{aligned}
W_E={}&\frac{1}{p+1}\Tr_{SO}X_1^{p+1}
+\frac{1}{p+1}\Tr_{USp}X_2^{p+1}\\
&+\Tr_{SO}(X_1X\widetilde X)
-\Tr_{USp}(X_2\widetilde X X).
\end{aligned}
\end{equation}
Equation~\eqref{eq:sosp-two-adjoint-rs} gives $(p+1)r=2$ and $r+2s=2$, so every term in~\eqref{eq:sosp-two-adjoint-electric-W} has $R$-charge two.

The elementary singlets in the magnetic theory correspond to the electric mesons
\begin{align}\label{eq:sosp-two-adjoint-mesons}
M_{11}^{(j)}&=Q_1X_1^jQ_1,
&j&=0,\ldots,2p-1,\nonumber\\
M_{22}^{(j)}&=Q_2X_2^jQ_2,
&j&=0,\ldots,2p-1,\\
M_{12}^{(\ell)}&=Q_1X_1^\ell XQ_2
\simeq Q_1XX_2^\ell Q_2,
&\ell&=0,\ldots,p-1.\nonumber
\end{align}
Their full $R$-charges are
\begin{equation*}
R(M_{11}^{(j)})=2R_{Q_1}+jr,\qquad
R(M_{22}^{(j)})=2R_{Q_2}+jr,\qquad
R(M_{12}^{(\ell)})=R_{Q_1}+R_{Q_2}+s+\ell r.
\end{equation*}
The diagonal mesons also carry a definite flavor parity. At the $SO$ node, $M_{11}^{(j)}$ is symmetric in flavor for even $j$ and antisymmetric for odd $j$. At the $USp$ node, the assignment is reversed: $M_{22}^{(j)}$ is antisymmetric for even $j$ and symmetric for odd $j$. Each $M_{12}^{(\ell)}$ is a bifundamental of $SU(N_1)\times SU(N_2)$.

The finite spectrum follows directly from the superpotential. Suppressing nonzero numerical coefficients, the electric F-term relations are
\begin{equation}\label{eq:sosp-two-adjoint-F-terms}
X_1^p+X\widetilde X=0,\qquad
X_2^p-\widetilde X X=0,\qquad
X_1X-XX_2=0,\qquad
\widetilde X X_1-X_2\widetilde X=0.
\end{equation}
The first three relations give
\begin{equation*}
X_1^pX=-X\widetilde X X=-XX_2^p,\qquad
X_1^pX=XX_2^p,
\end{equation*}
and hence $X_1^pX=XX_2^p=0$. Together with the last relation in~\eqref{eq:sosp-two-adjoint-F-terms}, this implies
\begin{equation*}
X_1^{2p}=0,\qquad X_2^{2p}=0.
\end{equation*}
The independent diagonal paths are therefore $1,X_i,\ldots,X_i^{2p-1}$, while an off-diagonal basis is $X_1^\ell X\simeq XX_2^\ell$ for $\ell=0,\ldots,p-1$. These paths give the meson towers in~\eqref{eq:sosp-two-adjoint-mesons}, which agree exactly with the large-$N$ index result in~\eqref{eq:sosp-two-adjoint-index-constraint}. For $p=1$, both adjoints are massive. Integrating them out gives the quartic $SO\times USp$ product-group duality, corresponding to the $k=0$ case discussed by Intriligator, Leigh, and Strassler~\cite{Intriligator:1995ax}. General odd $p$ gives the known two-adjoint family of~\cite{Ahn:1997gs}.

\subsubsection{Finite-rank SCI check}

As an additional finite-rank check, set all flavor fugacities to $1$, so the index depends only on $t$ and the spacetime fugacity $x$. For $p=1$, the expansion is evaluated at
\[
(N_1,N_2;N_{c1},N_{c2})=(3,7;7,6),
\qquad
(N_{c1}^{\rm d},N_{c2}^{\rm d})=(6,11).
\]
The beta-function and quark-pairing conditions give $(R_{Q_1},R_{Q_2})=(\frac{1}{3},\frac{5}{14})$ and $(R_{q_1},R_{q_2})=(\frac{1}{6},\frac{1}{7})$. The finite-rank electric and magnetic $SO$--$USp$ indices agree:
\begin{align}\label{eq:sosp-two-adjoint-sci-expansion}
\mathcal I_E^{SO/USp}=\mathcal I_M^{SO/USp}
={}&1+6t^{\frac{2}{3}}+21t^{\frac{5}{7}}+21t^{\frac{25}{21}}
+21t^{\frac{4}{3}}+126t^{\frac{29}{21}}+231t^{\frac{10}{7}}\nonumber\\
&+\left[3+6\left(x+x^{-1}\right)\right]t^{\frac{5}{3}}
+\left[28+21\left(x+x^{-1}\right)\right]t^{\frac{12}{7}}+\cdots.
\end{align}
The agreement through the displayed order provides a nontrivial finite-$N$ check of the $p=1$ $SO$--$USp$ duality.

\section{Conclusion and outlook}
\label{sec:conclusion}

We have developed a large-$N$ method for identifying and testing index-compatible candidate Seiberg dualities of $\mathcal{N}=1$ quiver gauge theories within a specified magnetic ansatz. The main results are:

\begin{enumerate}
\item A new closed-form expression, equation~\eqref{indwithfund}, for the large-$N$ index of the class of anomaly-free $\prod_{i=1}^n SU(N_{ci})$ quiver gauge theories described in Section~\ref{sec:largen}, with adjoint and oriented bifundamental matter and vector-like fundamental--antifundamental pairs. The formula is derived directly by a single fixed-mode complex Gaussian and is reproduced by sequential node-by-node integration in explicit low-node checks. It reduces to the known result~\cite{Dolan:2008qi} for $n=1$ and, when fundamental matter and gauge singlets are absent, to the $SU$ Gadde--Rastelli--Razamat--Yan formula including both the determinant and trace-exponential factors~\cite{Gadde:2010en}. Appendix~\ref{app:sosp} gives the corresponding real-Gaussian $SO/USp$ quiver formula used to determine meson spectra and parities.

\item Equality of the electric and magnetic large-$N$ indices gives the matrix constraint
\begin{equation*}
H(Q,t)-t^{\Delta+2}H(Q,t)^T
=\Bigl(\sum_{I^{ij}}t^{I^{ij}}\Bigr)_{n\times n}.
\end{equation*}
Within the magnetic ansatz, this equation is a necessary condition on the chiral spectrum and an algebraic test for large-$N$ index-compatible quiver-duality candidates. When $H(Q,t)$ is symmetric, the constraint takes the simpler form
\begin{equation*}
(1-t^{\Delta+2})H(Q,t)
=\Bigl(\sum_{I^{ij}}t^{I^{ij}}\Bigr)_{n\times n}.
\end{equation*}

\item The 't Hooft anomalies are checked at finite $N_c$ and $N_f$. A complementary baryon map is obtained when the required completeness and charge-pairing assumptions hold.

\item Explicit examples and checks: in the $SU$--$SU$ example, the $p=1$ limit recovers the Brodie--Hanany duality~\cite{Brodie:1997sz}, while the general-$p$ adjoint family reproduces the product-group duality of Brodie~\cite{Brodie:1996vx}. Section~\ref{sec:sosp-example} presents the corresponding two-adjoint $SO$--$USp$ family of Ahn, Oh, and Tatar~\cite{Ahn:1997gs}. Its $p=1$ limit reduces to the quartic $SO$--$USp$ product-group duality discussed by Intriligator, Leigh, and Strassler~\cite{Intriligator:1995ax}.
\end{enumerate}

Applying the matrix constraint~\eqref{eq:dual-matrix-form} recovers several known quiver dualities and yields a broader class of large-$N$ index-compatible candidates. Rather than dualizing one gauge factor at a time, the present analysis treats the entire gauge sector as a single unit and imposes one matrix constraint on all gauge nodes simultaneously. The off-diagonal entries of $H(Q,t)$ couple distinct gauge nodes and flavor channels directly. A systematic classification of these candidates, together with their finite-rank consistency conditions and operator maps, will be investigated elsewhere.

The Gaussian derivation suggests that the same determinant-and-inverse-kernel structure may extend to large-rank supersymmetric quiver matrix models beyond four dimensions. Monopole sectors in 3d, instantons in 5d, and vortex sectors in 2d would enter through the corresponding single-letter indices. Comparing global solutions of the constraint equation with sequences of ordinary single-node dualities may clarify which candidates factorize into known elementary moves and which are genuinely collective. For a quiver with $n$ gauge nodes and their associated flavor sectors, collections of $n-1,n-2,\ldots,2,1$ gauge nodes can be treated collectively as a single unit under Seiberg duality. Varying this collective unit provides a systematic way to generate further dual descriptions. Extensions to theories with strongly coupled matter may broaden the scope of index-based candidate searches~\cite{unpublished}.

\section*{Acknowledgments}

We thank Bohan Li for helpful discussions. We are especially grateful to Leonardo Santilli for inspiring discussions and for informing us of his unpublished work with Mohammed Akhond. The work of Dan Xie is supported by the National Key Research and Development Program of China (No.~2020YFA0713000).

\appendix
\section{Detailed derivation of the large-$N$ quiver index formula}
\label{app:derive-index}

This appendix derives the large-$N$ superconformal index formula~\eqref{indwithfund} for the class of $SU$ quivers described in Section~\ref{sec:largen}, with gauge group $\prod_{i=1}^n SU(N_{ci})$.

Conceptually, the derivation has three ingredients. The Haar measure supplies the universal quadratic term for trace variables. The Diaconis--Shahshahani moment formula identifies the Haar integrals of fixed finite trace polynomials with their Gaussian counterparts once the rank is sufficiently large. The remaining calculation is then an ordinary fixed-mode Gaussian integral, whose determinant gives the product factor in~\eqref{indwithfund} and whose source term gives the inverse-kernel contribution.

\subsection{Matrix integral, Haar measure, and trace variables}

\noindent\emph{Matrix-integral form.}

For a four-dimensional $\mathcal{N}=1$ gauge theory with gauge group $G$, the superconformal index is the protected trace over the Hilbert space obtained by radial quantization on $S^3\times\mathbb{R}$~\cite{Kinney:2005ej,Romelsberger:2005eg,Romelsberger:2007ec}:
\begin{equation*}
\mathcal{I}(t,x,\{a_i\})=\Tr_{\mathcal{H}_{S^3}}(-1)^F t^{\frac{2}{3}(E+j_2)}x^{2j_1}\prod_i a_i^{F_i},
\end{equation*}
where $E$ is the conformal dimension, $(j_1,j_2)$ are the $SU(2)\times SU(2)$ quantum numbers, and $F_i$ are flavor charges. This is the same fugacity convention as in the main text: a chiral scalar with $E=\frac{3}{2}R$ and $j_2=0$ contributes $t^R$. For a Lagrangian gauge theory, the index localizes to a matrix integral~\cite{Dolan:2008qi}:
\begin{equation}\label{eq:app-matrix-integral}
\mathcal{I}(t)=\int_G d\mu(U)\,\exp\Bigl(\sum_{m=1}^{\infty}\frac{1}{m}\,i(t^m,U^m)\Bigr),
\end{equation}
where $d\mu(U)$ is the normalized Haar measure on $G$ and $i(t,U)$ is the single-particle (single-letter) index, which can be read directly from the Lagrangian. In the matrix-integral formulas below, the spacetime fugacity and flavor fugacities are suppressed whenever they are spectators of the gauge integration.

For a quiver with gauge group $G=\prod_{i=1}^n SU(N_{ci})$, each factor contributes its own Haar measure and its own gauge fugacities. Writing $U_i\in SU(N_{ci})$ for the $i$-th gauge group, the full matrix integral is
\begin{equation*}
\mathcal{I}(t)=\int_{SU(N_{c1})}\!\!\!d\mu(U_1)\cdots\int_{SU(N_{cn})}\!\!\!d\mu(U_n)\,
\exp\Bigl(\sum_{m=1}^{\infty}\frac{1}{m}\,i(t^m,U_1^m,\dots,U_n^m)\Bigr),
\end{equation*}
where $i(t,U_1,\dots,U_n)$ is the same single-particle index as $i(t,z_1,\dots,z_n)$ in equation~\eqref{eq:general-single-particle}, written before passing to eigenvalue variables. More explicitly, for a single $SU(N)$ gauge node we choose a unitary matrix $V$ that diagonalizes $U$,
\begin{equation*}
U=V\,\diag(e^{i\theta_1},\dots,e^{i\theta_N})\,V^{-1}
\end{equation*}
and write
\begin{equation*}
z_\alpha=e^{i\theta_\alpha},\qquad \alpha=1,\dots,N,
\end{equation*}
with the $SU(N)$ constraint
\begin{equation*}
\prod_{\alpha=1}^N z_\alpha=1
\quad\Longleftrightarrow\quad
\sum_{\alpha=1}^N\theta_\alpha=0\quad(\mathrm{mod}\ 2\pi).
\end{equation*}
For the quiver notation used in the main text, $z_i$ denotes the full set of eigenvalue fugacities for the $i$-th gauge node,
\begin{equation*}
z_i=(z_{i,1},\dots,z_{i,N_{ci}}),\qquad
z_{i,\alpha}=e^{i\theta_{i,\alpha}},\qquad
\prod_{\alpha=1}^{N_{ci}}z_{i,\alpha}=1.
\end{equation*}

\medskip\noindent\emph{Eigenvalue Haar measure.}

With this diagonalization, the Haar integral of a class function can be written as~\cite{Mehta:2004}
\begin{equation*}
\int_{SU(N)}d\mu(U)\,f(U)=\frac{C_N}{N!}\int_{-\pi}^{\pi}\prod_{\alpha=1}^{N}\frac{d\theta_\alpha}{2\pi}\,\delta_{2\pi}\Bigl(\sum_\alpha\theta_\alpha\Bigr)\,\Delta(\theta)\,f(\theta_1,\dots,\theta_N),
\end{equation*}
where $C_N$ is fixed by $\int d\mu(U)=1$ and $\delta_{2\pi}(\phi)=\sum_{k\in\mathbb Z}\delta(\phi-2\pi k)$ is the periodic delta function imposing $\det U=1$. The Vandermonde determinant squared is
\begin{equation*}
\Delta(\theta)=\prod_{\alpha<\beta}\bigl|e^{i\theta_\alpha}-e^{i\theta_\beta}\bigr|^2
=\prod_{\alpha<\beta}4\sin^2\Bigl(\frac{\theta_\alpha-\theta_\beta}{2}\Bigr).
\end{equation*}
The integral over the diagonalizing matrix $V$ therefore factors out, leaving the eigenvalue measure.

\medskip\noindent\emph{Trace variables.}

A useful preliminary step is to Fourier-expand the logarithm of the Vandermonde determinant. Using the standard Fourier series result, valid for $0<|\theta|<2\pi$~\cite{Mehta:2004},
\begin{equation*}
\log\bigl|2\sin(\theta/2)\bigr|=-\sum_{m=1}^{\infty}\frac{\cos(m\theta)}{m},
\end{equation*}
the Weyl density becomes
\begin{align}\label{eq:app-vandermonde-fourier}
\log\Delta(\theta)&=2\sum_{\alpha<\beta}\log\bigl|2\sin\frac{\theta_\alpha-\theta_\beta}{2}\bigr|
=-2\sum_{\alpha<\beta}\sum_{m=1}^{\infty}\frac{\cos m(\theta_\alpha-\theta_\beta)}{m}\nonumber\\
&=-\sum_{m=1}^{\infty}\frac{1}{m}\sum_{\alpha\neq\beta}e^{im(\theta_\alpha-\theta_\beta)}
=-\sum_{m=1}^{\infty}\frac{1}{m}\bigl(p_m p_{-m}-N\bigr),
\end{align}
where we have introduced the \emph{trace variables}
\begin{equation*}
p_m(\theta)=\Tr U^m=\sum_{\alpha=1}^{N}e^{im\theta_\alpha},\qquad 
p_{-m}(\theta)=\Tr U^{-m}=\sum_{\alpha=1}^{N}e^{-im\theta_\alpha}=\overline{p_m(\theta)}.
\end{equation*}
The term proportional to $N$ in~\eqref{eq:app-vandermonde-fourier} is independent of the eigenangles and is absorbed into the normalization of the Haar measure. Thus, up to an overall normalization constant,
\begin{equation}\label{eq:app-haar-fourier}
d\mu(\theta)\propto\prod_{\alpha=1}^{N}d\theta_\alpha\,\delta_{2\pi}\Bigl(\sum_\alpha\theta_\alpha\Bigr)\,
\exp\Bigl[-\sum_{m=1}^{\infty}\frac{1}{m}p_m p_{-m}\Bigr].
\end{equation}

This expression reveals the quadratic nature of the Haar measure in the trace variables $p_m$. It is important to emphasize that \eqref{eq:app-haar-fourier} is \emph{not} yet a Gaussian integral: the $p_m$'s are highly nonlinear functions of the $N$ eigenangles, and at finite $N$ they are subject to numerous algebraic constraints.

\subsection{The Diaconis--Shahshahani theorem}

The key input that makes large-$N$ index computations tractable is the moment formula of Diaconis and Shahshahani~\cite{Diaconis:1994ap}. Here a moment means the normalized Haar integral of a monomial in the trace variables $p_{\pm m}(\theta)$ defined above. This is the form of their result that is directly needed for the index, because each fixed coefficient in the fugacity expansion is obtained by integrating a finite polynomial in finitely many trace variables.

For a fixed positive integer $K$, introduce complex variables $p_1,\dots,p_K$ with $p_{-m}:=\overline{p_m}$ and the finite-mode Gaussian measure
\begin{equation}\label{eq:app-ds-measure}
d\mu(p_1,\dots,p_K)=\prod_{m=1}^{K}\frac{d^2p_m}{\pi m}\,\exp\Bigl[-\frac{1}{m}p_m p_{-m}\Bigr],
\end{equation}
where $d^2p_m=d(\Re p_m)\,d(\Im p_m)$. Theorem~2 of~\cite{Diaconis:1994ap} says that the large-$N$ Haar moments of traces agree with the moments computed from this Gaussian measure. More explicitly, for fixed nonnegative integers $a_m,b_m$,
\begin{equation}\label{eq:app-ds-moments}
\resizebox{0.90\textwidth}{!}{$\displaystyle
\int_{U(N)}d\mu(U)\prod_{m=1}^K
\bigl(p_m^{(N)}(\theta)\bigr)^{a_m}\bigl(p_{-m}^{(N)}(\theta)\bigr)^{b_m}
=\prod_{m=1}^K\delta_{a_m,b_m}\,m^{a_m}a_m!
=\int d\mu(p_1,\dots,p_K)\prod_{m=1}^K p_m^{a_m}p_{-m}^{b_m}.
$}
\end{equation}
Here the equality holds once $N$ is larger than a number determined only by the fixed exponents $a_m,b_m$ and the largest mode $K$. For instance, it is enough to take $N$ larger than the total Fourier degree $\sum_{m=1}^K m(a_m+b_m)$. In~\eqref{eq:app-ds-moments}, the first expression is the Haar average over $U(N)$, the middle expression is its explicit Diaconis--Shahshahani value, and the last expression is the same moment evaluated with the Gaussian measure~\eqref{eq:app-ds-measure}. Therefore, by linearity, for any fixed finite polynomial $P_L(p_m,p_{-m})$, its Haar average at large $N$ equals the Gaussian average of the same polynomial:
\begin{equation*}
\int_{U(N)}d\mu(U)\,
P_L\bigl(p_m^{(N)},p_{-m}^{(N)}\bigr)
\longrightarrow
\int d\mu(p_1,\dots,p_K)\,
P_L(p_m,p_{-m}),
\end{equation*}
where the superscript $(N)$ on the left records the rank dependence of the trace variables $p_{\pm m}(\theta)$, while the measure on the right is~\eqref{eq:app-ds-measure}. This is the precise sense in which the fixed trace modes may be replaced by Gaussian variables inside a coefficient of the index.

Although the Diaconis--Shahshahani formula is stated for $U(N)$, it gives the same fixed-coefficient rule for the $SU(N)$ gauge integrals used here. Indeed, after expanding the plethystic exponential and choosing one coefficient, the integrand is a finite polynomial in finitely many trace modes. The constraint $\det U=1$ and possible epsilon-tensor invariants only affect terms whose degree grows with $N$. We therefore apply the Gaussian replacement to each fixed coefficient first, and only afterwards assemble these coefficients into the formal large-$N$ index.
\medskip

\subsection{Fixed-mode Gaussian for a single $SU(N)$ node}

The Diaconis--Shahshahani moment formula applied to the matrix integral~\eqref{eq:app-matrix-integral} gives the large-$N$ index for a single $SU(N_c)$ gauge group. For one $SU(N_c)$ node, the single-particle index has the form
\begin{equation*}
i(t,U)=f(t)\chi_{\mathrm{adj}}(U)+g(t)\chi_{\mathbf{N_c}}(U)+\overline{g}(t)\chi_{\overline{\mathbf{N_c}}}(U)+h(t),
\end{equation*}
where the first term includes the vector multiplet contribution. At the $m$-th plethystic mode these characters become
\[
\chi_{\mathrm{adj}}(U^m)=p_m p_{-m}-1,\qquad
\chi_{\mathbf{N_c}}(U^m)=p_m,\qquad
\chi_{\overline{\mathbf{N_c}}}(U^m)=p_{-m}.
\]

\paragraph{Finite-mode truncation.}
Fix a positive integer $K$ and truncate the plethystic sum in~\eqref{eq:app-matrix-integral} to $m\leq K$:
\begin{equation}\label{eq:app-truncated}
\mathcal{I}_N^{(K)}(t)=\int_{SU(N_c)}d\mu(U)\,
\exp\Bigl(\sum_{m=1}^{K}\frac{1}{m}\,i(t^m,U^m)\Bigr).
\end{equation}
After the denominators in the single-letter index are expanded as formal power series, $i(t,U)$ starts at a positive power of $t$. Let $d_{\min}$ be the smallest positive $t$-degree appearing in $i(t,U)$. Then the $m$-th plethystic mode $i(t^m,U^m)$ starts at order $t^{m d_{\min}}$. Therefore, to compute the index expansion up to and including order $t^L$, it is enough to keep the finitely many modes with $m d_{\min}\leq L$, or equivalently to choose the cutoff $K>L/d_{\min}$. Modes above this cutoff can only affect higher powers of $t$.

\paragraph{Trace variables in the plethystic exponent.}
At fixed mode $m$, the character expansion of the single-particle index can be written entirely in terms of $p_m$ and $p_{-m}$:
\begin{equation*}
i(t^m,U^m)=f_m\bigl(p_m p_{-m}-1\bigr)+g_m p_m+\overline{g}_m p_{-m}+h_m,
\end{equation*}
where we abbreviate $f_m:=f(t^m)$, $g_m:=g(t^m)$, $\overline{g}_m:=\overline{g}(t^m)$, $h_m:=h(t^m)$. The plethystic factor for mode $m$ is therefore
\begin{equation}\label{eq:app-plethystic-mode}
\exp\Bigl(\frac{1}{m}i(t^m,U^m)\Bigr)=
\exp\Bigl[\frac{1}{m}\Bigl(f_m(p_m p_{-m}-1)+g_m p_m+\overline{g}_m p_{-m}+h_m\Bigr)\Bigr].
\end{equation}

\paragraph{The fixed-mode Gaussian.}
At fixed cutoff $K$, expand any desired fugacity coefficient of~\eqref{eq:app-truncated}. The coefficient is a polynomial in the finite trace vector $(p_{\pm1},\dots,p_{\pm K})$, so the Diaconis--Shahshahani moment formula applies. Thus, after taking $N\to\infty$ at fixed $K$, the trace variables are integrated with the Gaussian density~\eqref{eq:app-ds-measure}. For the $m$-th mode, this Gaussian density contributes $-\frac{p_mp_{-m}}{m}$ in the exponent, while the plethystic exponent contributes~\eqref{eq:app-plethystic-mode}. The combined fixed-mode exponent is therefore
\begin{align*}
&-\frac{1}{m}p_m p_{-m}
+\frac{1}{m}\bigl[f_m(p_m p_{-m}-1)+g_m p_m+\overline{g}_m p_{-m}+h_m\bigr]\\
&\qquad
=\frac{1}{m}\Bigl[-(1-f_m)p_m p_{-m}
+g_m p_m+\overline{g}_m p_{-m}-f_m+h_m\Bigr].
\end{align*}

\paragraph{Evaluation of the Gaussian.}
Writing the Gaussian density inside the exponent, the integral for each mode $m$ becomes
\begin{equation*}
\int_{\mathbb{C}}\frac{d^2p_m}{\pi m}\,
\exp\Bigl[\frac{1}{m}\Bigl(-(1-f_m)p_m p_{-m}+g_m p_m+\overline{g}_m p_{-m}\Bigr)\Bigr]
\times\exp\Bigl[\frac{1}{m}\bigl(-f_m+h_m\bigr)\Bigr].
\end{equation*}
This is a standard complex Gaussian identity, applied as a formal power series around $1-f_m=1$:
\begin{align*}
\int_{\mathbb{C}}\frac{d^2p_m}{\pi m}\,
\exp\Bigl[-\frac{1-f_m}{m}|p_m|^2+\frac{1}{m}(g_m p_m+\overline{g}_m\,\overline{p_m})\Bigr]
&=\frac{1}{\pi m}\cdot\frac{\pi m}{1-f_m}\cdot\exp\Bigl(\frac{1}{m}\frac{g_m\overline{g}_m}{1-f_m}\Bigr)\\
&=\frac{1}{1-f_m}\exp\Bigl(\frac{1}{m}\frac{g_m\overline{g}_m}{1-f_m}\Bigr).
\end{align*}
Multiplication by the trace-variable independent factor $\exp[\frac{1}{m}(-f_m+h_m)]$ gives the contribution of mode $m$:
\begin{equation*}
\frac{1}{1-f_m}\exp\Bigl[\frac{1}{m}\Bigl(\frac{g_m\overline{g}_m}{1-f_m}-f_m+h_m\Bigr)\Bigr].
\end{equation*}

\paragraph{Removing the cutoff.}
For fixed $K$, the Diaconis--Shahshahani theorem is applied only to the finite set of modes $m=1,\dots,K$. In the Gaussian limit these modes are independent, and the exponent is a sum of separate $m$-mode terms. Therefore the truncated index is the finite product of the mode factors above, with $m=1,\dots,K$.

For a fixed order $t^L$, choosing $K>L/d_{\min}$ includes every mode that can contribute through that order. Increasing $K$ changes only higher powers of $t$. Hence every fixed-order expansion is determined by a finite product. Collecting the stabilized coefficients gives the formal infinite-product expression. All infinite products in this appendix are understood in this coefficient-wise sense, as in standard large-$N$ index computations~\cite{Dolan:2008qi}.
\begin{equation}\label{eq:app-single-final}
\boxed{\mathcal{I}(t)=\exp\Bigl(\sum_{m=1}^{\infty}\frac{1}{m}\Bigl[\frac{g(t^m)\overline{g}(t^m)}{1-f(t^m)}-f(t^m)+h(t^m)\Bigr]\Bigr)\prod_{m=1}^{\infty}\frac{1}{1-f(t^m)}.}
\end{equation}
\subsection{Extension to $n$-node quivers}

The single-node derivation extends to $\prod_{i=1}^n SU(N_{ci})$ quiver gauge theories by collecting the trace variables of all gauge nodes into one Gaussian.

\paragraph{Trace variables for several nodes.}
For each gauge node $i$, introduce trace variables
\begin{equation*}
p_{i,m}=\sum_{\alpha=1}^{N_{ci}}e^{im\theta_{i,\alpha}},\qquad
p_{i,-m}=\sum_{\alpha=1}^{N_{ci}}e^{-im\theta_{i,\alpha}}=\overline{p_{i,m}},
\qquad i=1,\dots,n,\;m=1,\dots,K.
\end{equation*}
Since the product Haar measure factorizes over the gauge nodes, the Diaconis--Shahshahani moment formula~\eqref{eq:app-ds-moments} can be applied independently to each $SU(N_{ci})$ factor. Thus, for any fixed polynomial in the finitely many variables $p_{i,\pm m}$, its large-$N$ Haar average is computed by the product Gaussian measure
\begin{equation*}
\prod_{i=1}^{n}\prod_{m=1}^{K}
\frac{d^2p_{i,m}}{\pi m}\,
\exp\Bigl[-\frac{1}{m}p_{i,m}p_{i,-m}\Bigr],
\qquad p_{i,-m}:=\overline{p_{i,m}}.
\end{equation*}

\paragraph{The single-particle index matrix.}
The matrix $\mathbf{i}(t)$ contains the gauge-sector single-letter data: diagonal entries come from vector multiplets and adjoint chirals, while off-diagonal entries come from bifundamental chirals. The vectors $\mathbf g$ and $\overline{\mathbf g}$ are flavor sources attached to the gauge nodes, $\tilde h$ contains gauge-singlet contributions, and the fixed-mode Gaussian kernel is $\mathbf K_m=\mathbf 1_n-\mathbf i(t^m)$.

The general single-particle index~\eqref{eq:general-single-particle} can be written compactly in terms of the trace variables as
\begin{equation}\label{eq:app-general-i}
i(t^m,U_1^m,\dots,U_n^m)=\sum_{i,j=1}^n p_{i,m}\,\mathbf{i}_{ij}(t^m)\,p_{j,-m}
+\sum_{i=1}^n\bigl(g_i(t^m)p_{i,m}+\overline{g}_i(t^m)p_{i,-m}\bigr)
-\tr\mathbf{i}(t^m)+\tilde h(t^m),
\end{equation}
where the $n\times n$ matrix $\mathbf{i}(t)$ has entries
\begin{equation*}
\mathbf{i}_{ij}(t)=
\begin{cases}
f_i(t), & i=j,\\
f_{ij}(t), & i\neq j.
\end{cases}
\end{equation*}
The diagonal terms $p_{i,m}\mathbf{i}_{ii}p_{i,-m}$ include the $-1$ from the adjoint character, which has been absorbed into the $-\tr\mathbf{i}(t^m)$ constant term together with the singlet subtractions.

\paragraph{The joint fixed-mode Gaussian.}
At fixed mode $m$, the combined exponent from the Diaconis--Shahshahani Gaussian density and the plethystic integrand is
\begin{equation*}
S_m=-\sum_{i=1}^n\frac{1}{m}p_{i,m}p_{i,-m}+\frac{1}{m}i(t^m,U_1^m,\dots,U_n^m).
\end{equation*}
Substitution of~\eqref{eq:app-general-i} organizes the exponent into quadratic and linear terms:
\begin{align*}
S_m&=\frac{1}{m}\Bigl[-\sum_{i,j=1}^n p_{i,m}\bigl(\delta_{ij}-\mathbf{i}_{ij}(t^m)\bigr)p_{j,-m}
+\sum_{i=1}^n\bigl(g_i(t^m)p_{i,m}+\overline{g}_i(t^m)p_{i,-m}\bigr)\Bigr]\\
&\quad+\frac{1}{m}\bigl[-\tr\mathbf{i}(t^m)+\tilde h(t^m)\bigr].
\end{align*}

Introduce the vector notation
\begin{equation*}
\mathbf{p}_m=\begin{pmatrix}p_{1,m}\\\vdots\\p_{n,m}\end{pmatrix},\qquad
\overline{\mathbf{p}}_m=\begin{pmatrix}p_{1,-m}\\\vdots\\p_{n,-m}\end{pmatrix},\qquad
\mathbf{g}_m=\begin{pmatrix}g_1(t^m)\\\vdots\\g_n(t^m)\end{pmatrix},\qquad
\overline{\mathbf{g}}_m=\begin{pmatrix}\overline{g}_1(t^m)\\\vdots\\\overline{g}_n(t^m)\end{pmatrix},
\end{equation*}
and the kernel matrix
\begin{equation}\label{eq:app-kernel}
\mathbf{K}_m:=\mathbf{1}_n-\mathbf{i}(t^m),
\end{equation}
where $\mathbf{1}_n$ is the $n\times n$ identity matrix. Its components are
\begin{equation*}
(\mathbf K_m)_{ij}=\delta_{ij}-\mathbf{i}_{ij}(t^m)=
\begin{cases}
1-f_i(t^m),& i=j,\\[2pt]
-f_{ij}(t^m),& i\neq j.
\end{cases}
\end{equation*}
Thus $\mathbf K_m$ is not an additional quiver matrix: it is precisely the fixed-mode value of the matrix $\mathbf 1-\mathbf i(t)$ appearing in the main formula~\eqref{indwithfund}. The identity term comes from the Haar measure of the $SU$ gauge nodes, while $-\mathbf i(t^m)$ subtracts the gauge-sector single-letter contributions. Equivalently, using~\eqref{eq:i-to-Mq},
\begin{equation*}
\mathbf K_m=\frac{(1-t^{2m})\mathbf 1_n-M_Q(t^m)+t^{2m}M_Q^T(t^{-m})}{(1-t^m x^m)(1-t^m x^{-m})}.
\end{equation*}
Consequently the determinant and source factors produced by the Gaussian are exactly $\det(\mathbf 1-\mathbf i(t^m))$ and $\overline{\mathbf g}(t^m)^T(\mathbf 1-\mathbf i(t^m))^{-1}\mathbf g(t^m)$.

With this notation, the $m$-th mode exponent becomes
\begin{equation*}
S_m=\frac{1}{m}\Bigl[-\mathbf{p}_m^T\mathbf{K}_m\overline{\mathbf{p}}_m+\mathbf{g}_m^T\mathbf{p}_m+\overline{\mathbf{g}}_m^T\overline{\mathbf{p}}_m\Bigr]
+\frac{1}{m}\bigl[-\tr\mathbf{i}(t^m)+\tilde h(t^m)\bigr].
\end{equation*}
\paragraph{Evaluation of the multivariate Gaussian.}
For fixed $m$, the Diaconis--Shahshahani theorem gives the measure $\prod_{i=1}^n d^2p_{i,m}/(\pi m)$. The $m$-th mode integral is
\begin{equation*}
\mathcal{I}_m=\int_{\mathbb{C}^n}\prod_{i=1}^n\frac{d^2p_{i,m}}{\pi m}\,
\exp\Bigl[\frac{1}{m}\Bigl(-\mathbf{p}_m^T\mathbf{K}_m\overline{\mathbf{p}}_m+\mathbf{g}_m^T\mathbf{p}_m+\overline{\mathbf{g}}_m^T\overline{\mathbf{p}}_m\Bigr)\Bigr].
\end{equation*}
The standard multivariate complex Gaussian identity is
\begin{equation*}
\int_{\mathbb{C}^n}\prod_{i=1}^n\frac{d^2z_i}{\pi}\,
\exp\bigl(-\mathbf{z}^T\mathbf{K}\overline{\mathbf{z}}+\mathbf{u}^T\mathbf{z}+\mathbf{v}^T\overline{\mathbf{z}}\bigr)
=\frac{1}{\det\mathbf{K}}\exp\bigl(\mathbf{v}^T\mathbf{K}^{-1}\mathbf{u}\bigr)
=\frac{1}{\det\mathbf{K}}\exp\bigl(\mathbf{u}^T\mathbf{K}^{-T}\mathbf{v}\bigr),
\end{equation*}
where $\overline{\mathbf z}$ is the complex conjugate of $\mathbf z$. Equivalently, in the Gaussian moment expansion, a $z_i$ can pair only with a $\overline z_j$, with the orientation recorded by $\mathbf K^{-1}$. The two displayed source forms are equal because the result is a scalar. We use this identity algebraically, as a formal expansion near $\mathbf K=\mathbf 1$. Since $\mathbf{K}_m=\mathbf{1}_n+O(t^\alpha)$ near $t=0$, both $\mathbf{K}_m^{-1}$ and $\det\mathbf{K}_m$ have well-defined formal power-series expansions.

The factors of $m$ come directly from the normalization of the $m$-th trace mode:
\begin{equation*}
\prod_{i=1}^n\frac{d^2p_{i,m}}{\pi m}=m^{-n}\prod_{i=1}^n\frac{d^2p_{i,m}}{\pi},
\end{equation*}
\begin{align*}
\frac{1}{m}\Bigl(-\mathbf p_m^T\mathbf K_m\overline{\mathbf p}_m
+\mathbf g_m^T\mathbf p_m+\overline{\mathbf g}_m^T\overline{\mathbf p}_m\Bigr)
&=-\mathbf p_m^T\frac{\mathbf K_m}{m}\overline{\mathbf p}_m
+\frac{\mathbf g_m^T}{m}\mathbf p_m+\frac{\overline{\mathbf g}_m^T}{m}\overline{\mathbf p}_m .
\end{align*}
Thus the Gaussian identity is applied with $\mathbf{K}=\frac{\mathbf{K}_m}{m}$, $\mathbf{u}=\frac{\mathbf{g}_m}{m}$, and $\mathbf{v}=\frac{\overline{\mathbf{g}}_m}{m}$, together with the prefactor $m^{-n}$ from the measure. This gives
\begin{align*}
\mathcal{I}_m&=\frac{1}{\det(\frac{\mathbf{K}_m}{m})}\cdot\frac{1}{m^n}\exp\Bigl(\frac{1}{m^2}\overline{\mathbf{g}}_m^T(m\mathbf{K}_m^{-1})\mathbf{g}_m\Bigr)\\
&=\frac{1}{\det\mathbf{K}_m}\exp\Bigl(\frac{1}{m}\overline{\mathbf{g}}_m^T\mathbf{K}_m^{-1}\mathbf{g}_m\Bigr).
\end{align*}
Multiplying by the constant factor $\exp[\frac{1}{m}(-\tr\mathbf{i}(t^m)+\tilde h(t^m))]$ gives the full contribution of the $m$-th mode. For fixed $K$, the truncated quiver index is the finite product of these contributions over $m\leq K$. As in the single-node case, any fixed order in the fugacity expansion only depends on finitely many modes. After $K$ is large enough, that coefficient no longer changes. The final large-$N$ formula is therefore the following formal infinite product, understood in this coefficient-by-coefficient sense:

\begin{equation}\label{eq:app-quiver-final}
\boxed{\mathcal{I}(t)=\exp\Bigl(\sum_{m=1}^{\infty}\frac{1}{m}\Bigl[\overline{\mathbf{g}}(t^m)^T\bigl(\mathbf{1}-\mathbf{i}(t^m)\bigr)^{-1}\mathbf{g}(t^m)-\tr\mathbf{i}(t^m)+\tilde h(t^m)\Bigr]\Bigr)
\prod_{m=1}^{\infty}\frac{1}{\det\bigl(\mathbf{1}-\mathbf{i}(t^m)\bigr)}.}
\end{equation}

The compact matrix form is invariant under permutations of the gauge nodes, consistent with the absence of a preferred integration order in the full fixed-mode Gaussian.

\subsection{Special cases and consistency checks}

\paragraph{Single gauge group.}
When $n=1$, the matrix $\mathbf{i}(t)$ reduces to the scalar $f(t)$. Then $\det(\mathbf{1}-\mathbf{i})=1-f$, $\tr(\mathbf{i})=f$, and $(\mathbf{1}-\mathbf{i})^{-1}=1/(1-f)$. Formula~\eqref{eq:app-quiver-final} becomes
\begin{equation*}
\mathcal{I}(t)=\exp\Bigl(\sum_{m=1}^{\infty}\frac{1}{m}\Bigl[\frac{g(t^m)\overline{g}(t^m)}{1-f(t^m)}-f(t^m)+h(t^m)\Bigr]\Bigr)\prod_{m=1}^{\infty}\frac{1}{1-f(t^m)},
\end{equation*}
recovering the single-node result~\eqref{eq:app-single-final}.

\paragraph{No fundamental flavors.}
When the quiver has no fundamental flavors, all $g_i$ and $\overline{g}_i$ vanish. Formula~\eqref{eq:app-quiver-final} then becomes
\begin{equation*}
\mathcal{I}(t)=
\exp\Bigl(\sum_{m=1}^{\infty}\frac{-\tr\mathbf{i}(t^m)+\tilde h(t^m)}{m}\Bigr)
\prod_{m=1}^{\infty}\frac{1}{\det\bigl(\mathbf{1}-\mathbf{i}(t^m)\bigr)}.
\end{equation*}
If gauge singlets are also absent, then $\tilde h=0$ and this is the $SU$ Gadde--Rastelli--Razamat--Yan formula, including the trace-exponential factor~\cite{Gadde:2010en}.

\paragraph{Sequential integration check.}
Sequential application of the single-node formula~\eqref{eq:app-single-final} gives an equivalent low-node check of~\eqref{eq:app-quiver-final}. At the $SU(N_{c1})$ integral, the trace characters $p_{i,m}$ of the remaining nodes enter the effective source $g_1^{\rm eff}$. The resulting action supplies the sources for the subsequent node integrals. Direct calculations for $n=2$ and $n=3$ reproduce~\eqref{eq:app-quiver-final}. For arbitrary $n$, the fixed-mode Gaussian gives the direct derivation, while sequential integration displays the same determinant and inverse-kernel structure iteratively.

\section{Large-$N$ index of $SO/USp$ quiver gauge theories}
\label{app:sosp}

This appendix develops the fixed-mode real-Gaussian formulation of the large-$N$ index for quivers with $SO$ and $USp$ gauge groups. Throughout, $USp(N_c)$ denotes the compact symplectic group with an even-dimensional fundamental representation of dimension $N_c$. Because the fundamental representations of $SO$ and $USp$ are self-conjugate, the fixed trace modes are real, and the large-rank matrix integral is a real rather than complex Gaussian. The single-node kernel extends to a general quiver by assembling all trace modes into one matrix Gaussian. We retain the determinant and normalized-Haar terms as universal components of the Gaussian formula. Flavor-neutral, trace-independent single-letter terms are denoted by $c_m$ and retained symbolically in the general formula, although they are omitted from the leading large-$N$ meson-spectrum comparison.

The quadratic part of the Haar measure determines the Gaussian kernel. Even-mode terms in the Haar measure and in rank-two matter characters shift the linear source, and their signs distinguish symmetric from antisymmetric flavor tensors. Within a specified magnetic ansatz, and for electric and magnetic theories with the same quadratic kernel, the index determines both the candidate meson spectrum and the $\mathbb Z_2$ parity of diagonal mesons.

\subsection{Simple $SO/USp$ node}

For $SO$ and $USp$ gauge groups, the fundamental representation is self-conjugate, so the gauge character is real. In analogy with the complex trace mode $p_m$ used for $SU$ groups in Appendix~\ref{app:derive-index}, we define the real angular Fourier mode
\begin{equation*}
q_m(\theta)=2\sum_{\alpha=1}^{r}\cos(m\theta_\alpha),
\end{equation*}
where $r$ is the rank: $N_c=2r$ for $SO(2r)$ and $USp(2r)$, while $N_c=2r+1$ for $SO(2r+1)$.
We write $\delta_{m\in2\mathbb Z}=1$ when $m$ is even and $\delta_{m\in2\mathbb Z}=0$ when $m$ is odd.

As in the $SU$ case, the input needed for the index is a Haar/Gaussian moment identity. Theorem~4 for $O(N)$ and Theorem~6, equation~(4.1), for the compact symplectic group give the required moment formulas for the full fundamental traces $T_m(U)=\Tr U^m$~\cite{Diaconis:1994ap}. In the angular variables used in the index,
\begin{equation*}
T_m(U)=q_m(U)+\kappa_G,
\qquad
\kappa_G=\begin{cases}
0,&G=SO(2r)\ \text{or}\ USp(2r),\\
1,&G=SO(2r+1).
\end{cases}
\end{equation*}
The corresponding Gaussian mean of $q_m$ is
\begin{equation*}
\mu_{G,m}=\begin{cases}
\delta_{m\in2\mathbb Z},&G=SO(2r),\\
-\delta_{m\in2\mathbb Z},&G=USp(2r),\\
-1+\delta_{m\in2\mathbb Z},&G=SO(2r+1).
\end{cases}
\end{equation*}
For $G_r=SO(2r)$, $USp(2r)$, or $SO(2r+1)$ and any fixed polynomial $P$, the moment identity used in the index is
\begin{equation}\label{eq:sosp-ds-moments}
\begin{aligned}
&\int_{G_r}d\mu_{G_r}(U)\,
P\bigl(q_1(U),\dots,q_K(U)\bigr)\\
&\qquad=
\int_{\mathbb R^K}\prod_{m=1}^{K}
\left[
\frac{dq_m}{\sqrt{2\pi m}}\,
\exp\Bigl(-\frac{(q_m-\mu_{G,m})^2}{2m}\Bigr)
\right]P(q_1,\dots,q_K),
\end{aligned}
\end{equation}
once the rank is sufficiently large compared with the weighted degree of $P$. On the right-hand side, the $q_m$ are independent real Gaussian integration variables with mean $\mu_{G,m}$ and variance $m$. Equation~\eqref{eq:sosp-ds-moments} equates polynomial averages and does not assert a pointwise equality between the trace of a particular matrix and a Gaussian variable. For a monomial $P=\prod_{m=1}^K q_m^{r_m}$, the weighted degree is $d=\sum_{m=1}^Kmr_m$. The Diaconis--Shahshahani formulas are exact at finite rank when $N_c\geq d$ for $O(N_c)$ and when $r\geq d$ for $USp(N_c)=USp(2r)$, equivalently $N_c\geq2d$. The corresponding $SO(N_c)$ identity follows from the $O(N_c)$ formula when $N_c>d$, as explained next.

Theorem~4 is stated for $O(N)$ rather than $SO(N)$. The passage to $SO(N)$ uses a separate fixed-degree argument. A trace monomial of weighted degree $d$ involves at most $d$ fundamental indices. The invariant tensors common to $O(N)$ and $SO(N)$ are generated by Kronecker-delta contractions, while an $SO(N)$-specific epsilon contraction requires $N$ indices. Therefore, when $N>d$, no epsilon contraction can occur, and the $O(N)$ and $SO(N)$ Haar moments of the trace monomial agree. Since every fixed fugacity coefficient of the index is a finite polynomial in finitely many trace modes, equation~\eqref{eq:sosp-ds-moments} gives the coefficient-wise Gaussian replacement required below.

As in Appendix~\ref{app:derive-index}, all Gaussian expressions below are understood coefficient by coefficient in the non-gauge fugacities. At any fixed fugacity order, only finitely many trace modes and finite powers of them occur, so the moment identity applies to a finite polynomial. The kernel inverses and determinant square roots are consequently interpreted as formal power series around the identity kernel, with the square-root branch chosen to have constant term one.

The normalized pure-Haar Gaussian factor for mode $m$ is therefore
\begin{equation*}
\frac{dq_m}{\sqrt{2\pi m}}
\exp\Bigl[-\frac{(q_m-\mu_{G,m})^2}{2m}\Bigr]
=\frac{dq_m}{\sqrt{2\pi m}}
\exp\Bigl[\frac{1}{m}\Bigl(-\frac{q_m^2}{2}
+\mu_{G,m}q_m-\frac{\mu_{G,m}^2}{2}\Bigr)\Bigr].
\end{equation*}
Thus the Haar measure supplies the universal quadratic kernel and the group-dependent linear source. The even-mode contribution is positive for $SO$ and negative for $USp$, while $\kappa_G$ records the fixed eigenvalue of an odd-orthogonal fundamental representation.

Having established the Gaussian form of the Haar measure, we now include the vector and matter single-letter contributions. At plethystic mode $m$, define
\begin{align*}
D_m&=(1-t^m x^m)(1-t^m x^{-m}),&
\gamma_R^{(m)}&=\frac{t^{mR}-t^{m(2-R)}}{D_m},\\
v_m&=f_{\rm vec}(t^m)=\frac{2t^{2m}-t^m(x^m+x^{-m})}{D_m}.
\end{align*}
Here $\gamma_R^{(m)}$ is the coefficient multiplying the character of a chiral multiplet of $R$-charge $R$ at mode $m$, while $v_m$ is the vector-multiplet coefficient. For a rank-two chiral $X_A$, we write the former as $\gamma_{R_{X_A}}^{(m)}$.

The rank-two matter signs follow directly from the character identities
\begin{equation*}
\chi_{\operatorname{Sym}^2}(U^m)=\frac{T_m^2+T_{2m}}{2},
\qquad
\chi_{\wedge^2}(U^m)=\frac{T_m^2-T_{2m}}{2}.
\end{equation*}
Consequently, for a rank-two chiral multiplet $X_A$ with $R(X_A)=R_{X_A}$, the non-constant part of its character in the $q$ variables is
\begin{equation*}
\frac{q_m^2+\tau_A q_{2m}}{2}+\kappa_G q_m,\qquad
\tau_A=\begin{cases}+1,&X_A\ \text{symmetric},\\-1,&X_A\ \text{antisymmetric}.
\end{cases}
\end{equation*}
Thus a symmetric tensor has $\tau_A=+1$, while an antisymmetric tensor has $\tau_A=-1$. The adjoint representation is antisymmetric for $SO$ and symmetric for $USp$, so $\tau_{\rm adj}=-1$ for $SO$ and $\tau_{\rm adj}=+1$ for $USp$. After multiplying by the single-letter coefficient $\gamma_{R_{X_A}}^{(m)}$, the $q_m^2$ term contributes to the quadratic coefficient $k_m$, the $q_{2m}$ term contributes to the even-mode source coefficient $e_m$, and the $\kappa_G q_m$ term is included in the shifted source defined below.

For the simple node, let $g_m$ denote the total coefficient of the term linear in $q_m$ in the fixed-mode exponent, so the fundamental matter contribution is $g_mq_m/m$. The term $h_m$ contains flavor-dependent contributions that are independent of $q_m$, including gauge singlets. For $SO(2r+1)$, the fixed eigenvalue gives the additional trace-independent contribution $\kappa_Gg_m$ to $h_m$. Flavor-neutral, trace-independent single-letter contributions of order $O(N^0)$ are collected in $c_m$. More explicitly, after writing $T_m=q_m+\kappa_G$, a full symmetric or antisymmetric rank-two chiral $X_A$ contributes
\begin{equation*}
\gamma_{R_{X_A}}^{(m)}\frac{\kappa_G(1+\tau_A)}{2}
\end{equation*}
to $c_m$. For a quiver, a bifundamental chiral $X_{ij}$ between nodes $i$ and $j$ contributes
\begin{equation*}
\gamma_{R_{X_{ij}}}^{(m)}\kappa_{G_i}\kappa_{G_j}.
\end{equation*}
The complete $c_m$ is obtained by summing these terms over all rank-two and bifundamental chirals. Flavor-neutral elementary singlets and any singlet subtraction required for a traceless tensor representation must also be included. The normalized-Haar term $-\mu_{G,m}^2/2$ is displayed separately and is not part of $c_m$. For the two-adjoint $SO$--$USp$ theory in Section~\ref{sec:sosp-example}, these contributions vanish, so $c_m=0$. The flavor-refined source for an individual quiver node is given in~\eqref{eq:sosp-general-refined-source} below.

To derive the fixed-mode effective action, we now collect the normalized Haar, vector, rank-two matter, fundamental, and trace-independent contributions before performing the Gaussian integral. The integrand is
\begin{equation}\label{eq:sosp-fixed-mode-integrand}
\begin{aligned}
\prod_{\ell=1}^{\infty}\frac{dq_\ell}{\sqrt{2\pi\ell}}
\exp\Bigg\{\sum_{\ell=1}^{\infty}\frac{1}{\ell}\Bigg[&
-\frac{1}{2}q_\ell^2
+\mu_{G,\ell}q_\ell
-\frac{\mu_{G,\ell}^2}{2}
\\
&+v_\ell\left(\frac{1}{2}q_\ell^2+\kappa_Gq_\ell
+\frac{\tau_{\rm adj}}{2}q_{2\ell}\right)
\\
&+\sum_A\gamma_{R_{X_A}}^{(\ell)}
\left(\frac{1}{2}q_\ell^2+\kappa_Gq_\ell
+\frac{\tau_A}{2}q_{2\ell}\right)
+g_\ell q_\ell+h_\ell+c_\ell\Bigg]\Bigg\}.
\end{aligned}
\end{equation}
In~\eqref{eq:sosp-fixed-mode-integrand}, the first line comes from the Haar measure. The second line is the vector-multiplet contribution, with $\tau_{\rm adj}=-1$ for $SO$ and $\tau_{\rm adj}=+1$ for $USp$. The third line sums over the rank-two chirals $X_A$. Finally, $g_\ell q_\ell$ is the ordinary flavor-refined source. For an odd orthogonal node, the fixed eigenvalue also produces the trace-independent flavor term $\kappa_Gg_\ell$, which is included in $h_\ell$ together with elementary gauge singlets.

We now organize the exponent according to the final integration variable $q_m$. Contributions to $q_m$ arise directly from terms evaluated at plethystic mode $m$. When $m$ is even, additional linear contributions arise from terms of the form $q_{2\ell}$ evaluated at the lower mode $\ell=m/2$.

The terms quadratic in $q_m$ are
\begin{equation*}
-\frac{1}{2}q_m^2+\frac{v_m}{2}q_m^2
+\frac{1}{2}\sum_A\gamma_{R_{X_A}}^{(m)}q_m^2
=-\frac{k_m}{2}q_m^2,
\qquad
k_m=1-v_m-\sum_A\gamma_{R_{X_A}}^{(m)}.
\end{equation*}
The terms linear in $q_m$ at the same plethystic mode consist of the fundamental source $g_mq_m$, the parity-independent Haar contribution $-\kappa_Gq_m$, and the terms proportional to $\kappa_Gq_m$ from the vector multiplet and rank-two chirals. Their combined coefficient is
\begin{equation*}
g_m-\kappa_G+\kappa_Gv_m
+\kappa_G\sum_A\gamma_{R_{X_A}}^{(m)}
=g_m-\kappa_Gk_m.
\end{equation*}
It remains to collect the terms containing $q_{2\ell}$. They contribute to the coefficient of $q_m$ only when $m$ is even. Setting $m=2\ell$, the vector and rank-two chiral terms generated at mode $\ell$ become
\begin{align*}
\frac{1}{\ell}\left[\frac{\tau_{\rm adj}v_\ell}{2}
+\frac{1}{2}\sum_A\tau_A\gamma_{R_{X_A}}^{(\ell)}\right]q_{2\ell}
=\frac{1}{m}\left[\tau_{\rm adj}v_{\frac m2}
+\sum_A\tau_A\gamma_{R_{X_A}}^{(\frac m2)}\right]q_m.
\end{align*}
The Haar mean supplies one further even-mode term, namely $+q_m/m$ for $SO$ and $-q_m/m$ for $USp$. Combining it with the expression above gives
\begin{equation*}
\frac{1}{m}e_{\frac m2}q_m,
\qquad
e_m=\begin{cases}
1-v_m+\displaystyle\sum_A\tau_A\gamma_{R_{X_A}}^{(m)},&G=SO(2r)\ \text{or}\ SO(2r+1),\\[4pt]
-(1-v_m)+\displaystyle\sum_A\tau_A\gamma_{R_{X_A}}^{(m)},&G=USp(2r).
\end{cases}
\end{equation*}
Here the argument of $e$ records the lower mode at which the $q_{2\ell}$ term was generated. Combining the same-mode and lower-mode contributions, the complete coefficient of $q_m$ is
\begin{equation*}
\widehat g_m^G
=g_m-\kappa_Gk_m
+\delta_{m\in2\mathbb Z}\,e_{\frac m2}.
\end{equation*}
For odd $m$, the last term is absent. For even $m=2\ell$, $e_{m/2}=e_\ell$ is evaluated at the lower plethystic mode $\ell$.

The complete retained exponent is therefore
\begin{equation*}
\sum_{m=1}^{\infty}\frac{1}{m}
\left[-\frac{k_m}{2}q_m^2+\widehat g_m^Gq_m+h_m+c_m
-\frac{\mu_{G,m}^2}{2}\right].
\end{equation*}
Defining the fixed-mode integrand as $\exp[-S_G^{\rm eff}]$ gives
\begin{align*}
S_G^{\rm eff}&=\sum_{m=1}^{\infty}\frac{1}{m}
\left[\frac{k_m}{2}q_m^2-(\widehat g_m^G)q_m-h_m-c_m
+\frac{\mu_{G,m}^2}{2}\right].
\end{align*}
Evaluating this Gaussian mode by mode gives the simple-node result
\begin{equation}\label{eq:sosp-simple-index}
\boxed{
\mathcal I_G(t)
=\exp\Biggl\{\sum_{m=1}^{\infty}\frac{1}{m}
\left[\frac{(\widehat g_m^G)^2}{2k_m}+h_m+c_m
-\frac{\mu_{G,m}^2}{2}\right]\Biggr\}
\prod_{m=1}^{\infty}\frac{1}{\sqrt{k_m}}.}
\end{equation}
The product in~\eqref{eq:sosp-simple-index} is the determinant factor for the real Gaussian, while the term $-\mu_{G,m}^2/2$ restores the normalized-Haar factor. To check the normalization, remove all vector and matter letters. Then the ordinary matter source vanishes, $g_m=0$, and
\begin{align*}
v_m&=0,& \gamma_{R_{X_A}}^{(m)}&=0,& k_m&=1,& h_m&=c_m=0,\\
e_m&=\begin{cases}
+1,&G=SO(2r)\ \text{or}\ SO(2r+1),\\
-1,&G=USp(2r).
\end{cases}
\end{align*}
The definition of the shifted source therefore gives
\begin{equation*}
\widehat g_m^G
=-\kappa_G+\delta_{m\in2\mathbb Z}e_{\frac m2}
=\mu_{G,m}.
\end{equation*}
Hence~\eqref{eq:sosp-simple-index} reduces to $\mathcal I_G=1$, as required by $\int_Gd\mu_G(U)=1$.

\subsection{$SO/USp$ quiver formula}
\label{app:sosp-quiver-formula}

For a quiver with multiple $SO/USp$ gauge nodes, the single-node trace modes combine into a real matrix Gaussian.

Let a fundamental chiral $Q_i$ of $R$-charge $R_{Q_i}$ and flavor group $SU(N_i)$ be attached to gauge node $i$. At mode $m$, its contribution linear in the gauge trace $q_{i,m}$ is $g_{i,m}^{(Q_i)}q_{i,m}/m$, where
\begin{equation}\label{eq:sosp-general-refined-source}
g_{i,m}^{(Q_i)}
=\frac{t^{mR_{Q_i}}P_{i,m}-t^{m(2-R_{Q_i})}\overline P_{i,m}}{D_m},
\qquad
P_{i,m}=\Tr_{\mathbf N_i}y_i^m,
\quad
\overline P_{i,m}=P_{i,-m}.
\end{equation}
For a flavor antifundamental, $P_{i,m}$ and $\overline P_{i,m}$ are interchanged. The source $g_{i,m}$ is the sum of the fundamental and antifundamental contributions attached to node $i$. If node $i$ is $SO(2r_i+1)$, its fixed eigenvalue also contributes the trace-independent term $\kappa_{G_i}g_{i,m}$ to $h_m$.

For a quiver with self-conjugate $SO/USp$ gauge nodes, let $q_{i,m}$ be the real trace mode of node $i$, and collect them into $\mathbf q_m=(q_{1,m},\dots,q_{n,m})^T$. We also define the vector of pure-Haar means
\begin{equation*}
\boldsymbol\mu_m=(\mu_{G_1,m},\dots,\mu_{G_n,m})^T.
\end{equation*}
The retained fixed-mode effective action is
\begin{equation*}
S_{\rm quiver}^{\rm eff}=\sum_{m=1}^{\infty}\frac{1}{m}\Bigl[
\frac{1}{2}\mathbf q_m^T K_m\mathbf q_m
-\widehat{\mathbf g}_m^T\mathbf q_m-h_m-c_m
+\frac{1}{2}\boldsymbol\mu_m^T\boldsymbol\mu_m\Bigr],
\end{equation*}
so $K_m$ is the symmetric kernel of the real Gaussian and $\widehat{\mathbf g}_m$ is the shifted source vector. Its entries are determined directly by the quiver data:
\begin{equation*}
(K_m)_{ij}=
\begin{cases}
\displaystyle 1-v_{i,m}
-\sum_{\substack{X_A\,\text{rank-two chiral}\\\text{at node }i}}
\gamma_{R_{X_A}}^{(m)},&i=j,\\[12pt]
\displaystyle -\sum_{\substack{X\,\text{bifundamental chiral}\\\text{between nodes }i\text{ and }j}}
\gamma_{R_X}^{(m)},&i\neq j.
\end{cases}
\end{equation*}
Here $v_{i,m}$ is the vector-multiplet coefficient at node $i$. The first sum includes every rank-two chiral attached to that node, while the second includes every independent bifundamental chiral joining the two nodes. The off-diagonal sign follows from the quadratic exponent: for $i\neq j$, the term $-\frac{1}{2}\mathbf q_m^T K_m\mathbf q_m$ contains $-(K_m)_{ij}q_{i,m}q_{j,m}$, whereas the bifundamental single-letter index contributes $+\sum_X\gamma_{R_X}^{(m)}q_{i,m}q_{j,m}$.

This is the $SO/USp$ analogue of the $SU$-quiver kernel $\mathbf K_m=\mathbf 1_n-\mathbf i(t^m)$ in~\eqref{eq:app-kernel}. In both cases, the kernel is the coefficient matrix of the quadratic terms in the fixed-mode Gaussian. For $SU$ nodes the trace modes are complex variables $p_{i,m},p_{i,-m}$, and in the convention of Appendix~\ref{app:derive-index} the quadratic term has the form $-\mathbf p_m^T\mathbf K_m\overline{\mathbf p}_m$. With the sources written as $\mathbf g_m^T\mathbf p_m+\overline{\mathbf g}_m^T\overline{\mathbf p}_m$, this gives the source factor $\overline{\mathbf g}_m^T\mathbf K_m^{-1}\mathbf g_m=\mathbf g_m^T\mathbf K_m^{-T}\overline{\mathbf g}_m$ for a nonsymmetric quiver kernel. For $SO/USp$ nodes the fundamental character is self-conjugate, so the trace modes $q_{i,m}$ are real. The quadratic term is $-\frac{1}{2}\mathbf q_m^T K_m\mathbf q_m$, and the real Gaussian produces $\det(K_m)^{-\frac{1}{2}}$. Thus $K_m$ plays the same role as $\mathbf 1-\mathbf i(t^m)$, but for the real self-conjugate trace variables. The additional even-mode and odd-orthogonal Haar contributions are linear in the fixed modes and are therefore absorbed into the shifted source $\widehat{\mathbf g}_m$, rather than into the kernel $K_m$.

It is useful to separate the even-mode source coefficient
\begin{equation*}
e_{i,m}=\begin{cases}
1-v_{i,m}+\displaystyle\sum_{A\in i}\tau_A\gamma_{R_{X_A}}^{(m)},
&G_i=SO(2r_i)\ \text{or}\ SO(2r_i+1),\\[4pt]
-(1-v_{i,m})+\displaystyle\sum_{A\in i}\tau_A\gamma_{R_{X_A}}^{(m)},
&G_i=USp(2r_i).
\end{cases}
\end{equation*}
The shifted source vector $\widehat{\mathbf g}_m$ has components
\begin{equation*}
\widehat g_{i,m}=g_{i,m}-\sum_j\kappa_{G_j}(K_m)_{ji}+\delta_{m\in2\mathbb Z}\,e_{i,\frac{m}{2}}.
\end{equation*}
Here $g_{i,m}$ is the flavor-refined source at node $i$, as in~\eqref{eq:sosp-general-refined-source}. The sum over $j$ includes $j=i$ and is nonzero only when at least one node is odd orthogonal, where $\kappa_{G_j}=1$. The last term is present only at even modes: when $m=2\ell$, the coefficient $e_{i,\ell}$ supplies the linear source inherited from the $q_{2\ell}$ terms. If every orthogonal node is even, all $\kappa$'s vanish and the only Haar shift is the even-mode term. For an odd orthogonal node, both terms must be retained. The term $h_m$ denotes all flavor-dependent terms that are independent of the gauge trace modes, including elementary singlets and fixed-eigenvalue contributions.

Applying the real Gaussian identity mode by mode gives the $SO/USp$ quiver formula
\begin{equation}\label{eq:sosp-quiver-index}
\boxed{
\mathcal I_{SO/USp}(t)
=\exp\Biggl\{\sum_{m=1}^{\infty}\frac{1}{m}
\left[\frac{1}{2}\widehat{\mathbf g}_m^T K_m^{-1}
\widehat{\mathbf g}_m+h_m+c_m
-\frac{1}{2}\boldsymbol\mu_m^T\boldsymbol\mu_m\right]\Biggr\}
\prod_{m=1}^{\infty}\frac{1}{\sqrt{\det K_m}}.}
\end{equation}
To organize the large-$N$ comparison, we take the flavor ranks $N_i$ to scale with the gauge ranks, while the number of nodes and the plethystic mode $m$ remain fixed. Then $P_{i,m}=\Tr_{\mathbf N_i}y_i^m$ and the ordinary flavor source $\mathbf g_m$ are of order $O(N)$, whereas $K_m$, $\boldsymbol\mu_m$, and the shift $\widehat{\mathbf g}_m-\mathbf g_m$ are of order $O(N^0)$. The source-square term separates as
\begin{align}\label{eq:sosp-source-order-decomposition}
\frac{1}{2}\widehat{\mathbf g}_m^T K_m^{-1}\widehat{\mathbf g}_m
={}&\underbrace{\frac{1}{2}\mathbf g_m^T K_m^{-1}\mathbf g_m}_{O(N^2)}
+\underbrace{\mathbf g_m^T K_m^{-1}(\widehat{\mathbf g}_m-\mathbf g_m)}_{O(N)}\nonumber\\
&+\underbrace{\frac{1}{2}(\widehat{\mathbf g}_m-\mathbf g_m)^T
K_m^{-1}(\widehat{\mathbf g}_m-\mathbf g_m)}_{O(N^0)}.
\end{align}
The $O(N^0)$ term in~\eqref{eq:sosp-source-order-decomposition}, the determinant, the normalized-Haar subtraction, and $c_m$ are retained in the full formula~\eqref{eq:sosp-quiver-index}, but omitted from the meson-spectrum comparison below.

\subsection{Implications for duality}

We now summarize the leading $O(N^2)$ information for a proposed electric--magnetic pair with the same Gaussian kernel. The $O(N)$ terms further distinguish the flavor parities of the diagonal mesons, but we do not analyze them in detail here.

\paragraph{\texorpdfstring{$O(N^2)$ contribution.}{O(N2) contribution.}}
We define the denominator-cleared quadratic kernel $\mathcal K(t)$ by
\begin{equation*}
K_m=\frac{\mathcal K(t^m)}{D_m},
\qquad
D_m=(1-t^mx^m)(1-t^mx^{-m}).
\end{equation*}
Using the quiver-data definition of $K_m$ above, its entries are
\begin{equation*}
\mathcal K_{ij}(t)=
\begin{cases}
\displaystyle 1-t^2
-\sum_{\substack{X_A\,\text{rank-two chiral}\\\text{at node }i}}
\left(t^{R_{X_A}}-t^{2-R_{X_A}}\right),&i=j,\\[12pt]
\displaystyle -\sum_{\substack{X\,\text{bifundamental chiral}\\\text{between nodes }i\text{ and }j}}
\left(t^{R_X}-t^{2-R_X}\right),&i\neq j.
\end{cases}
\end{equation*}
Comparing with the $SU$-quiver definition~\eqref{eq:Hilbert-series-def}, $\mathcal K(t)$ corresponds to $H(Q,t)^{-1}$, while $\mathcal K(t)^{-1}$ is the $SO/USp$ analogue of the matrix Hilbert series. At the level of the quadratic kernel, the adjoint chirals in the diagonal entries of the $SU$ quiver matrix are replaced by rank-two chirals, and all independent bifundamental chirals joining two self-conjugate gauge nodes contribute to the symmetric off-diagonal entries. Symmetric and antisymmetric rank-two chirals have the same quadratic contribution, so $\mathcal K(t)$ does not distinguish them. Their difference is encoded in the $O(N)$ even-mode source terms, whose detailed analysis is left for future work.

The ordinary source square determines the total multiplicities and $R$-charges of the mesons through the coefficients of $P_{i,m}P_{j,m}$. For $i\neq j$, these coefficients describe bifundamental flavor mesons. For $i=j$, the characters $\operatorname{Sym}^2(\mathbf N_i)$ and $\wedge^2(\mathbf N_i)$ have the same leading term $P_{i,m}^2/2$, so this order determines only the total number of diagonal mesons. Repeating the $SU$-quiver source matching of Section~\ref{sec:duality} and using the quark-pairing condition~\eqref{eq:quark-pairing} gives
\begin{equation}\label{eq:sosp-meson-matrix-constraint}
\boxed{
(1-t^{\Delta+2})\mathcal K(t)^{-1}
=\begin{pmatrix}
\displaystyle\sum_{I^{11}}t^{I^{11}} & \cdots & \displaystyle\sum_{I^{1n}}t^{I^{1n}}\\[8pt]
\vdots & \ddots & \vdots\\[4pt]
\displaystyle\sum_{I^{n1}}t^{I^{n1}} & \cdots & \displaystyle\sum_{I^{nn}}t^{I^{nn}}
\end{pmatrix}.}
\end{equation}
The factor of $1/2$ in the diagonal Gaussian source square matches the leading term $P_{i,m}^2/2$ in both rank-two flavor characters. Thus no additional factor is required, and each matrix entry gives the total number and $R$-charges of the mesons in that flavor channel.

\bibliographystyle{JHEP}
\bibliography{ref1}
\end{document}